\documentclass[a4paper,11pt]{article}
\pdfoutput=1 

\usepackage{jheppub} 

\usepackage[T1]{fontenc} 
\usepackage{multirow}
\usepackage{diagbox}
\usepackage{rotating}
\usepackage{lineno}
\usepackage{tabularx}
\usepackage{setspace}

\usepackage{lineno}

\title{\boldmath The $Higgs\to b\bar{b}, c\bar{c}, gg$ measurement at CEPC }

\author[a,b,1]{Yongfeng Zhu,\note{Also at Some University.}}
\author[a,b,2]{Hanhua Cui,\note{Also at Some University.}}
\author[a,b,3]{Manqi Ruan\note{Corresponding author.}}

\affiliation[a]{Institute of High Energy Physics, Chinese Academy of Sciences,\\Beijing 100049, China}
\affiliation[b]{University of Chinese Academy of Sciences,\\Beijing 100049, China}

\emailAdd{ruanmq@ihep.ac.cn}

\abstract{
Accurately measuring the properties of the Higgs boson is one of the core physics objectives of the Circular Electron Positron Collider (CEPC). 
As a Higgs factory, the CEPC is expected to operate at a centre-of-mass energy of $240\,GeV$, deliver an integrated luminosity of $5.6\,ab^{-1}$, and produce one million Higgs bosons according to the CEPC Conceptual Design Report (CDR).
Combining measurements of the $\ell^+\ell^-H$, $\nu\bar{\nu} H$, and $q\bar{q}H$ channels, we conclude that the signal strength of $H\to b\bar{b}/c\bar{c}/gg$ can be measured with a relative accuracy (relative statistical uncertainty only) of 0.27\%/4.03\%/1.56\%.
Extrapolating to the recently released TDR operating parameters corresponding to the integrated luminosity of $20\,ab^{-1}$, the relative accuracy of $H\to b\bar{b}/c\bar{c}/gg$ signal strength is 0.14\%/2.13\%/0.82\% (relative statistical uncertainty only).
We analyze the dependence of the expected accuracies on the critical detector performances: Color Singlet Identification (CSI) for the $q\bar{q}H$ channel and flavor tagging for both $\nu\bar{\nu} H$ and $q\bar{q}H$ channels.  
Compared to the baseline CEPC detector performance, ideal flavor tagging can increase the $H\to b\bar{b}/c\bar{c}/gg$ signal strength accuracy by 2\%/63\%/13\% in the $\nu\bar{\nu} H$ channel and 35\%/122\%/181\% in the $q\bar{q}H$ channel.
A strong dependence between the CSI performance and anticipated accuracies in $q\bar{q}H$ channel is identified.
The relevant systematic uncertainties are also discussed in this paper.

}

\arxivnumber{2203.01469}

\begin{document} 
\maketitle
\flushbottom

\section{Introduction}
\label{sec:intro}
Lepton-Higgs factories~\cite{FCCCDR, ILC:2019gyn, Robson:2018enq, CEPCStudyGroup:2018ghi}, could precisely determine the properties of the Higgs boson, provide crucial information on top of the HL-LHC~\cite{HLLHC} and search for New Physics signatures beyond the Standard Model (SM). 
Intensive studies of the physics potential of various future facilities have been conducted~\cite{ESBook}, leading to the conclusion that "an electron-positron Higgs factory is the highest priority for the next collider" ~\cite{ES}.
Many electron-positron Higgs factories are proposed,
including the International Linear Collider (ILC)~\cite{ILC:2019gyn}, the Compact Linear $e^+e^-$ Collider (CLIC)~\cite{Robson:2018enq},  the Future Circular Collider $e^+e^-$ (FCC-ee)~\cite{FCC:2018evy}, and the Circular Electron Positron Collider (CEPC)~\cite{CEPCStudyGroup:2018ghi}.

\begin{figure}[tbp]
\centering 
\includegraphics[width=0.55\textwidth]{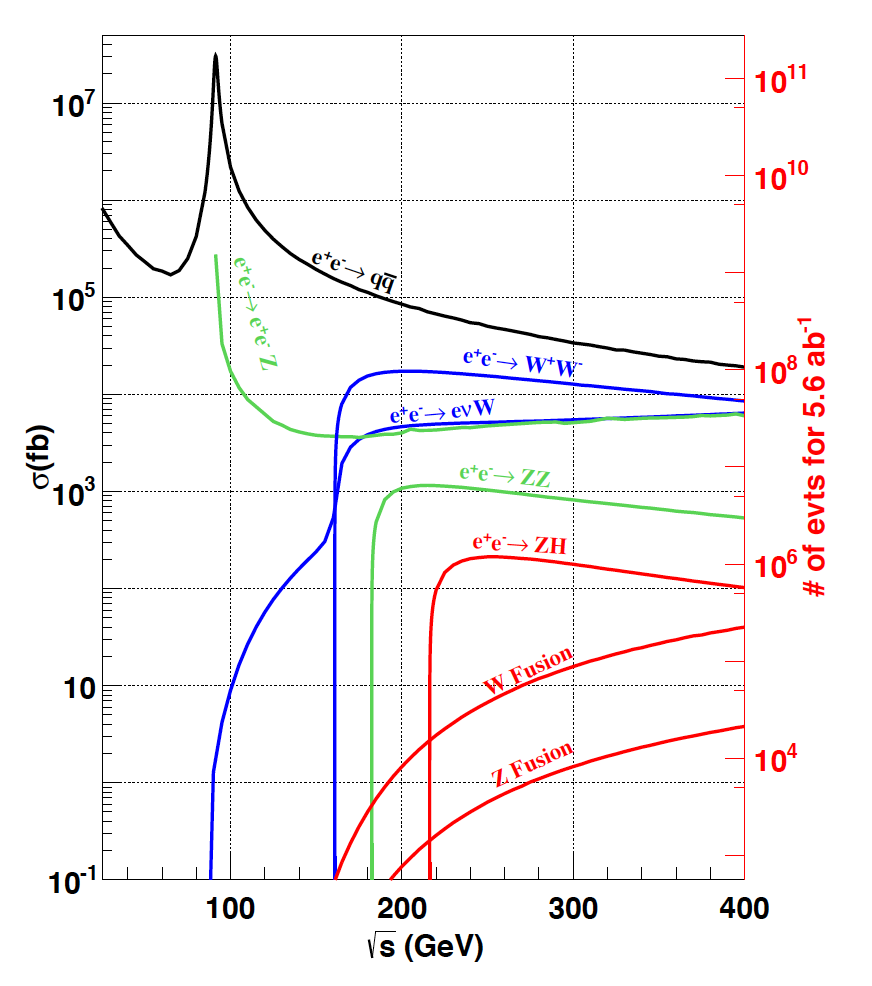}
\caption{\label{fig:x}The cross section for an unpolarized $e^{+}e^{-}$ collision~\cite{An:2018dwb}, the right side shows the expected number of events at the nominal parameters of the CEPC Higgs runs at $240\,GeV$ centre-of-mass energy.}
\end{figure}

\begin{figure}[tbp]
\centering 
\includegraphics[width=0.8\textwidth]{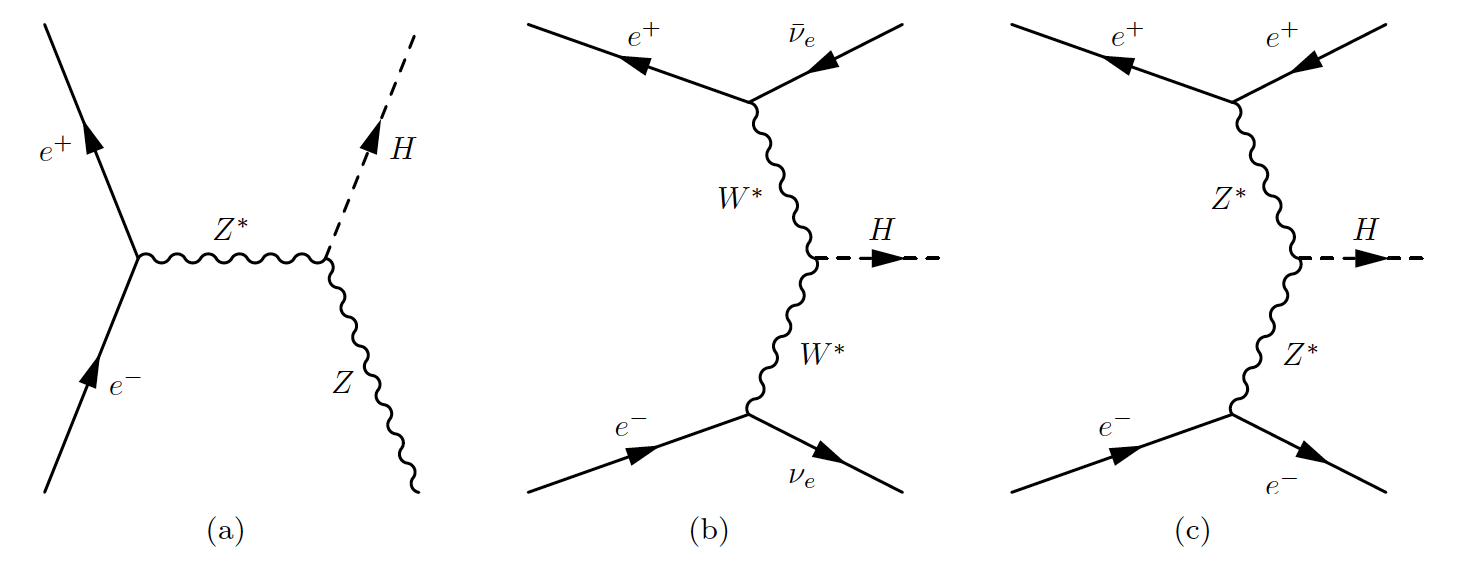}
\caption{\label{fig:1} The Feynman diagrams of the Higgs boson production processes in electron-positron collisions~\cite{An:2018dwb}, (a)$e^+e^-\to ZH$, (b)$e^+e^-\to \nu_e\bar{\nu}_eH$ and (c)$e^+e^-\to e^+e^-H$.}
\end{figure}

The CEPC is designed with a circumference of 100~km and two interaction points~\cite{CEPCAcc}.
It can operate at multiple centre-of-mass energies, including $240\,GeV$ as a Higgs factory, $160\,GeV$ for the $W^+W^-$ threshold scan,
and $91\,GeV$ as a Z factory. 
The main SM processes and corresponding cross sections are shown in figure~\ref{fig:x}.
It also has the potential to increase its centre-of-mass energy to $360\,GeV$ for top-quark pair production.
In the future, it can be upgraded to a proton-proton collider to directly search for new physics signals at a centre-of-mass energy of about $100\,TeV$, which is an order of magnitude higher than the LHC.
When operating at $240\,GeV$, the CEPC could produce Higgs bosons by the processes of Higgs-strahlung (ZH), WW fusion ($e^+e^-\to \nu_e\bar{\nu}_e H$), and ZZ fusion ($e^+e^-\to e^+e^- H$), with more than 96\% of Higgs bosons produced by the ZH process.
Their Feynman diagrams are shown in figure~\ref{fig:1}.

Measuring the branching fractions of the $H\to b\bar{b}/c\bar{c}/gg$ decays is one of the core CEPC physics objectives. 
This paper evaluates the statistical accuracies achievable for these measurements.
According to the particles generated in association with the Higgs boson, the analysis channels are classified into three categories: $\ell^+\ell^-H$, $\nu\bar{\nu} H$, and $q\bar{q}H$.
The expected performance in the $\ell^+\ell^-H$ channel is analyzed in ref.~\cite{Bai:2019qwd}.
This paper focuses on the analysis and detector performance optimization studies in the $\nu\bar{\nu} H$ and $q\bar{q}H$ channels and combines the results from all three channels.

This paper is organized into five sections. Section 2 introduces the detector, the software, and the simulated data samples used in this analysis.
Section 3 presents the analyzes in the $\nu\bar{\nu} H$ and $q\bar{q}H$ channels, and combines the results from all three channels.
Section 4 analyzes the dependence of objective accuracies on critical detector performances,
including flavor tagging performance and Color Singlet Identification (CSI)~\cite{CSI}.
Section 5 discusses various systematic uncertainties and possible strategies for controlling them.
The conclusions are summarized in the last section.

\section{Detector, softwares and data samples}
\label{sec:basic}

\begin{figure}[tbp]
\centering 
\includegraphics[width=0.55\textwidth]{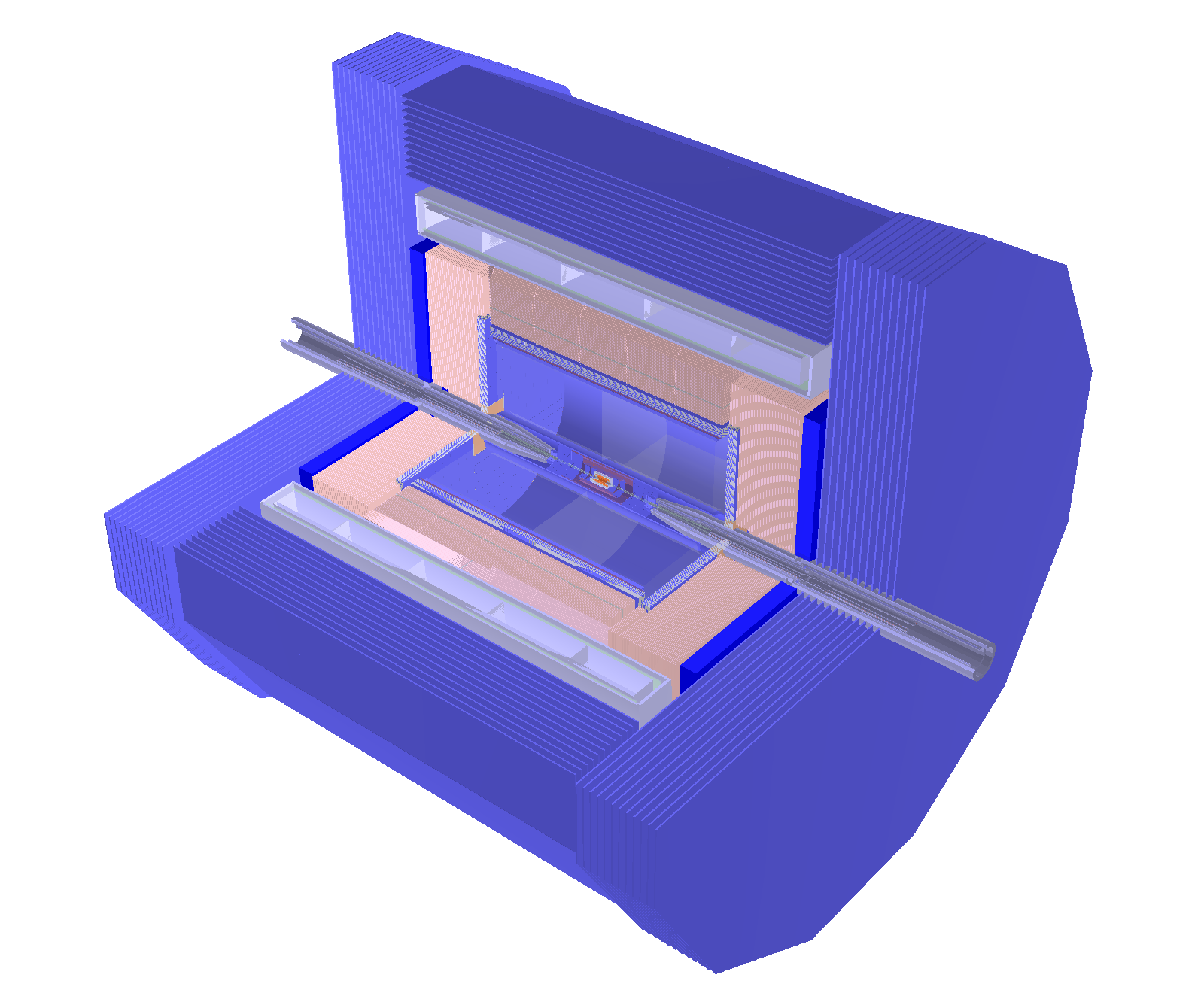}
\caption{\label{det}The CEPC baseline detector~\cite{CEPCStudyGroup:2018ghi}. From inner to outer, the detector is composed of a silicon pixel vertex detector, a silicon inner tracker, a TPC, a silicon external tracker, an ECAL, an HCAL, a solenoid of 3 Tesla and a return yoke embedded with a muon detector. In the forward regions, five pairs of silicon tracking disks are installed to enlarge the tracking acceptance.}
\end{figure}

The CEPC uses a Particle Flow Oriented (PFO) detector as its baseline detector~\cite{CEPCStudyGroup:2018ghi}.
This detector reconstructs and identifies all visible particles in the final state and measures their energy and momentum in the most-suited subdetector systems.
From inner to outer, this detector is composed of a silicon pixel vertex detector, a silicon inner tracker, a Time Projection Chamber (TPC) surrounded by a silicon external tracker, a silicon-tungsten sampling Electromagnetic Calorimeter (ECAL), a steel-Glass Resistive Plate Chambers (GRPC) sampling Hadronic Calorimeter (HCAL), a 3 Tesla superconducting solenoid, and a flux return yoke embedded a muon detector.
The structure of the CEPC detector is shown in figure~\ref{det}.
Precise measurement of the $H\to b\bar{b}/c\bar{c}/gg$ branching fractions require good jet flavor tagging performance, which is highly dependent on the CEPC vertex detector.
The vertex detector is described in subsection~\ref{sec:OptFLV}, which focuses on the optimization of flavor tagging performance.

\begin{figure}[tbp]
\centering
\includegraphics[width=0.65\textwidth]{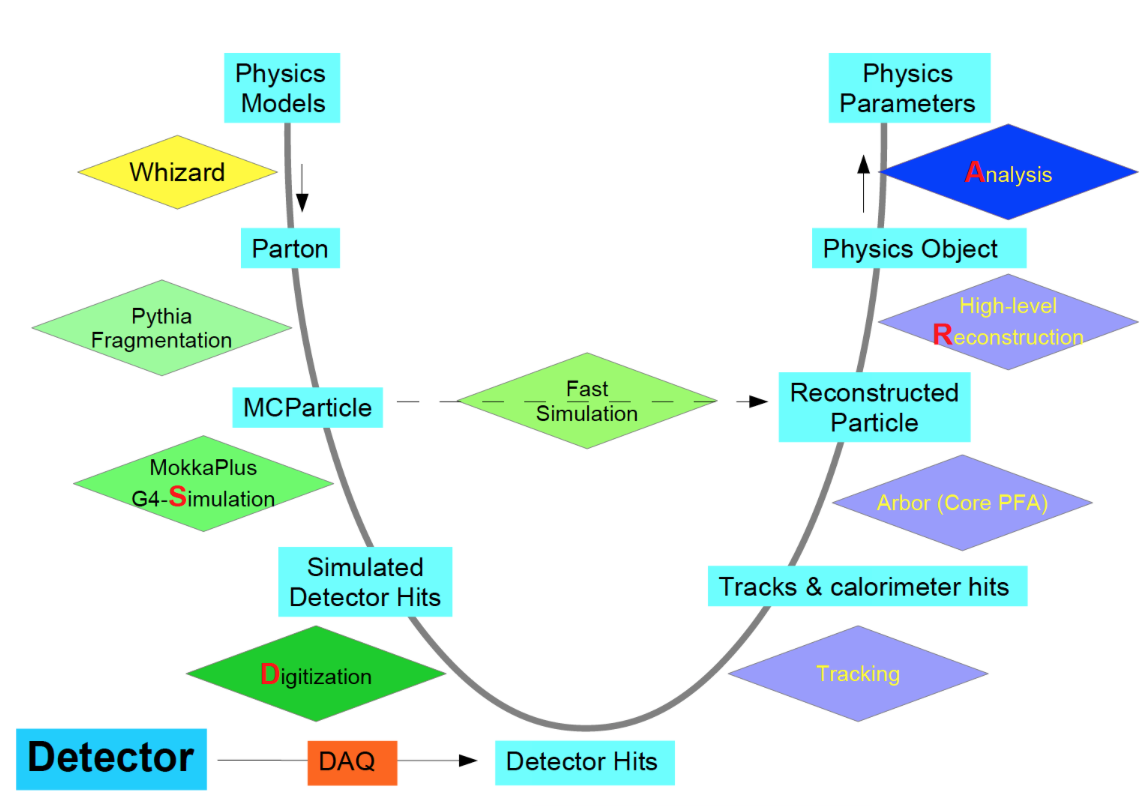}
\caption{\label{fig:softwarechain}The information flow of the CEPC software chain~\cite{Zhu:2018ift}.}
\end{figure}

A baseline reconstruction software chain (see figure~\ref{fig:softwarechain}) was developed to estimate the physics potential based on the simulation and reconstruction of physics objects with realistic detector effects.
The data flow of the CEPC baseline software starts with the event generators of WHIZARD~\cite{whizard} and PYTHIA~6.4~\cite{pythia}, unless explicitly stated.
The detector geometry is implemented in MokkaPlus~\cite{mokka}, a GEANT4-based simulation framework.
MokkaPlus calculates the energy deposition in the sensitive volumes and produces simulated hits.
The reconstruction modules include tracking, particle flow, and high-level reconstruction algorithms.
The tracker hits are reconstructed into tracks based on CLUPATRA~\cite{track}.
The Particle Flow algorithm, ARBOR~\cite{arbor}, reads the reconstructed tracks and the calorimeter hits to build reconstructed particles.
The physics objects, including electrons, muons, taus, missing energy, jets, etc., are reconstructed from the reconstructed particles.

The CEPC detector is expected to record at least one million Higgs boson events and about one billion SM background events, see figure~\ref{fig:x}. 
We classify the SM backgrounds into several categories, including two-fermion and four-fermion processes.
The two-fermion processes include the $q\bar{q}$, Bhabha, $\mu^+\mu^-$, and $\tau^+\tau^-$ processes, while four-fermion processes include the single-Z, single-W, ZZ, WW, and mixed processes.
The mixed processes are used to properly model the interference between intermediate processes, i.e. the 4-quark final state $u\bar{u}d\bar{d}$, which can be generated from both ZZ($Z\to u\bar{u}$, $Z\to d\bar{d}$) and $W^+W^-$($W^+\to u\bar{d}$, $W^-\to \bar{u}d$) processes.

The samples used in this paper were fully simulated with the CEPC baseline detector concept and reconstructed with its baseline software. 
The following analyses and referred factors are based on the integrated luminosity of $5.6\,ab^{-1}$.
We simulate 217,000/566,000 $\nu\bar{\nu}H$(including WW fusion)/$q\bar{q}H$ events, corresponding to 84\%/74\% of the statistics predicted by SM. 
We also simulated 47 million SM background events, including all major SM processes. 
The ratio between the statistics of the simulated events and the prediction of SM is then referred to as the scaling factor. 
To maximize the efficiency of limited computational resources, four-fermion backgrounds are assigned larger scaling factors (20\% - 83\%), 
while the scaling factors for two-fermion backgrounds range from 0.23\% to 2.8\%, 
as the overall statistics of two-fermion backgrounds are huge, but relatively easy to distinguish from the signal event. 
The $\gamma\gamma\to hadrons$ process is also included and is generated with PYTHIA 8.3~\cite{Bierlich:2022pfr}.
Two photons are produced by the incoming electrons and their interaction yields hadrons.
The detailed description of this process can be found in~\cite{gamma, Helenius:2017aqz}.
Because $\gamma\gamma\to hadrons$ background can be easily separated from the signal events, we simulate 1.36 million samples corresponding to 0.28\% of its total statistics. 
The appendix~\ref{appendix} describes the detailed information on the samples, including the process, cross section, expected event number, simulated event number, and scaling factor.

\section{Measurement of the relative statistical uncertainties}
The objective observables of our analyzes are the number of $\nu\bar{\nu}H$ and $q\bar{q}H$ events, with the Higgs decaying into jets fragmented from quarks or gluons. 
The analysis processes are generally divided into two steps.
The first step is to distinguish the Higgs-to-two-jets signal events from the background events.
The second step is to separate different Higgs decay modes based on the flavor tagging information. 
In the ref.~\cite{Bai:2019qwd}, the analysis process in the $\ell^+\ell^- H$ channel is described in detail.
We briefly summarize the analysis process in subsection~\ref{sec:llH}.
The analysis processes of $\nu\bar{\nu} H$ and $q\bar{q}H$ are described in subsections~\ref{sec:vvH} and \ref{sec:qqH}, respectively.

\subsection{\texorpdfstring{$\ell^+\ell^-H$}.}
\label{sec:llH}
This analysis is based on a centre-of-mass energy of $250\,GeV$ and an integrated luminosity of $5\,ab^{-1}$, which is slightly different from the normal setting with a centre-of-mass energy of $240\,GeV$ and an integrated luminosity of $5.6\,ab^{-1}$.
The signal events have two isolated leptons, $e^+e^-$ or $\mu^+\mu^-$, mostly from the Z-boson decay in the Higgs-strahlung process.
The invariant mass and recoil mass of these two leptons should be close to the Z boson and the Higgs boson, respectively.
The signal events also have two jets generated from the Higgs boson decay. 
Jet kinematics, i.e. the invariant mass and angle of the two jets, is used to improve the separation performance between signal and background.

After event selection, a template fitting method is used to determine the component fractions of the $H\to b\bar{b}$, $H\to c\bar{c}$, and $H\to gg$ processes. 
The relative accuracy of the $H\to b\bar{b}/c\bar{c}/gg$ signal strength, corresponding to the centre-of-mass energy of $250\,GeV$ and an integrated luminosity of $5\,ab^{-1}$, is 1.1\%/10.5\%/5.4\% in the $\mu^+\mu^-H$ channel and 1.6\%/14.7\%/10.5\% in the $e^+e^-H$ channel.
The relative accuracy in the $\mu^+ \mu^- H$ channel is better than that in the $e^+e^- H$ channel because, first, the background in the $e^+e^- H$ channel is significantly larger than that in the $\mu^+ \mu^- H$ channel due to the single-Z processes, and second, the momentum resolution for $\mu^{\pm}$ is better than that for $e^{\pm}$.
Extrapolating to the CEPC nominal settings under the assumption that the signal efficiencies and background suppression rates are the same at the two centre-of-mass energies, the $H\to b\bar{b}/c\bar{c}/gg$ signal strength accuracy is 1.57\%/14.43\%/10.31\% in the $e^+e^-H$ channel and 1.06\%/10.16\%/5.23\% in the $\mu^+\mu^-H$ channel.

\subsection{ \texorpdfstring{$\nu\bar{\nu} H$}. }
\label{sec:vvH}
This subsection describes the measurement of the branching fractions of $H\to b\bar{b}/c\bar{c}/gg$ in the $\nu\bar{\nu} H$ channel.
The first step focuses on the separation of Higgs-to-two-jets signal events from the entire sample corresponding to SM prediction with a cut-based event selection and the TMVA tool~\cite{TMVA}.
The cut variables are designed according to the characteristics of the signal and background, which are described below, and the cut flow is summarized in table~\ref{tab:3}. 
The $\gamma\gamma$ is the abbreviation for the $\gamma\gamma\to hadrons$ process.

\begin{table}[tbp]
\centering
\begin{scriptsize}
\begin{tabular}{|ccccccccccc|}
\hline
                                 & $\nu\bar{\nu} H q\bar{q}/gg$       &2f                & SW             &SZ               &WW             &ZZ              &Mixed          &ZH     & $\gamma\gamma$   &$\frac{\sqrt{S+B}}{S}$(\%)      \\
total                          &  178890                         &$8.01E8$      &$1.95E7$    &$9.07E6$    & $5.08E7$    &$6.39E6$   &$2.18E7$    &$961606$     &  $4.91E8$   & 20.92         \\
\hline
recoilMass (GeV)                & \multirow{2}{*}{157822}                         &\multirow{2}{*}{$5.11E7$}     &\multirow{2}{*}{$2.17E6$}  &\multirow{2}{*}{$1.38E6$}    &\multirow{2}{*}{$4.78E6$}    &\multirow{2}{*}{$1.30E6$}   &\multirow{2}{*}{$1.08E6$}      &\multirow{2}{*}{$74991$}     &   \multirow{2}{*}{$2.69E7$}    & \multirow{2}{*}{5.98}    \\
$\in(74, 131)$               &&&&&&&&&& \\
$visEn$ (GeV)                    & \multirow{2}{*}{142918}                         &\multirow{2}{*}{$2.37E7$}     &\multirow{2}{*}{$1.35E6$}  &\multirow{2}{*}{$8.81E5$}     &\multirow{2}{*}{$3.60E6$}   &\multirow{2}{*}{$1.03E6$}   &\multirow{2}{*}{$6.29E5$}     &\multirow{2}{*}{$50989$}     &   \multirow{2}{*}{$1.31E7$}    & \multirow{2}{*}{4.67}    \\
$\in(109, 143)$              &&&&&&&&&& \\
$leadLepEn$ (GeV)            &\multirow{2}{*}{141926}                         &\multirow{2}{*}{$2.08E7$}     &\multirow{2}{*}{$3.65E5$}    &\multirow{2}{*}{$7.24E5$}     &\multirow{2}{*}{$2.81E6$}    &\multirow{2}{*}{$9.72E5$}    &\multirow{2}{*}{$1.34E5$}     &\multirow{2}{*}{$46963$}       &   \multirow{2}{*}{$1.31E7$}    & \multirow{2}{*}{4.41}    \\
$\in(0, 42)$               &&&&&&&&&& \\
$multiplicity$                    &\multirow{2}{*}{139545}                          &\multirow{2}{*}{$1.66E7$}       &\multirow{2}{*}{$2.36E5$}    &\multirow{2}{*}{$5.24E5$}     &\multirow{2}{*}{$2.62E6$}   &\multirow{2}{*}{$9.07E5$ }       &\multirow{2}{*}{$4977$}       &\multirow{2}{*}{$42751$}  &  \multirow{2}{*}{$1.24E7$}  &\multirow{2}{*}{4.15}     \\
$\in(40, 130)$               &&&&&&&&&& \\
$leadNeuEn$ (GeV)        & \multirow{2}{*}{138653}                          &\multirow{2}{*}{$1.46E7$}       &\multirow{2}{*}{$2.24E5$}       &\multirow{2}{*}{$4.72E5$}     &\multirow{2}{*}{$2.49E6$}     &\multirow{2}{*}{$8.69E5$}        &\multirow{2}{*}{$4552$}          &\multirow{2}{*}{$42303$}    &  \multirow{2}{*}{$1.10E7$}   &\multirow{2}{*}{3.94}      \\
$\in(0, 41)$              &&&&&&&&&& \\
$Pt$ (GeV)                        & \multirow{2}{*}{121212}                        &\multirow{2}{*}{$248715$}       &\multirow{2}{*}{$1.56E5$}      &\multirow{2}{*}{$2.48E5$}     &\multirow{2}{*}{$1.51E6$}     &\multirow{2}{*}{$4.31E5$}        &\multirow{2}{*}{$999$}            &\multirow{2}{*}{$35453$}     &  \multirow{2}{*}{1437}   &\multirow{2}{*}{1.37}    \\
$\in(20, 60)$             &&&&&&&&&& \\
$Pl$ (GeV)                            & \multirow{2}{*}{118109}                        &\multirow{2}{*}{$52784$}          &\multirow{2}{*}{$1.05E5$}        &\multirow{2}{*}{74936}          &\multirow{2}{*}{$7.30E5$}     & \multirow{2}{*}{$1.13E5$}         & \multirow{2}{*}{$847$}           & \multirow{2}{*}{34279}     &  \multirow{2}{*}{1078}   & \multirow{2}{*}{0.94}        \\
$\in(0, 50)$            &&&&&&&&&& \\
-log10(Y23)                         &\multirow{2}{*}{96156}                          &\multirow{2}{*}{$40861$}           &\multirow{2}{*}{$26088$}         &\multirow{2}{*}{60349}          &\multirow{2}{*}{$2.25E5$}      &\multirow{2}{*}{82560}              &\multirow{2}{*}{$640$}            &\multirow{2}{*}{10691}       &    \multirow{2}{*}{1078}    & \multirow{2}{*}{0.76}    \\
$\in(3.375, +\infty)$             &&&&&&&&&& \\
InvMass (GeV)                 &\multirow{2}{*}{71758}                           &\multirow{2}{*}{22200}              & \multirow{2}{*}{11059}           &\multirow{2}{*}{6308}            &\multirow{2}{*}{77912}           &\multirow{2}{*}{13680}               & \multirow{2}{*}{248}               & \multirow{2}{*}{6915}        &    \multirow{2}{*}{359}        & \multirow{2}{*}{0.64}           \\
$\in(110, 134)$             &&&&&&&&&& \\
\hline
BDT                      & \multirow{2}{*}{60887}                           &\multirow{2}{*}{9140}              &\multirow{2}{*}{266}               &\multirow{2}{*}{2521}            &\multirow{2}{*}{3761}             &\multirow{2}{*}{3916}                &\multirow{2}{*}{58}                   & \multirow{2}{*}{1897}         &       \multirow{2}{*}{$0^*$}      & \multirow{2}{*}{0.47}           \\
$\in(-0.02, 1)$          &&&&&&&&&& \\
\hline
\end{tabular}
\end{scriptsize}
\caption{\label{tab:3} The event selection of $\nu\bar{\nu} H(H\to q\bar{q}/gg)$ is based on the integrated luminosity of $5.6\,ab^{-1}$. The $\gamma\gamma$ is the abbreviation for $\gamma\gamma\to hadrons$ process. The symbol $0^*$ represents that the number of $\gamma\gamma\to hadrons$ events is less than $3.09/0.0028$ at a confidence level of 95\% according to  Feldman-Cousins approach in the case of observing a zero event, where 3.09 is quoted from ref.~\cite{Feldman:1997qc} and 0.0028 is the scaling factor of the $\gamma\gamma\to hadrons$ process.}
\end{table}

\begin{itemize}
\item Most of the $\nu\bar{\nu}H$ events are from ZH process with $Z\to \nu\bar{\nu}$, while the WW-fusion contributes about 13\% (ignore the interference when calculating the contribution of WW fusion to $\nu\bar{\nu}H$).
The signal events have only the jets from the Higgs boson decay. 
Therefore, the signal should have a recoil mass peak at the mass of the Z boson.
However, the SM process of $\nu\bar{\nu} Z(Z\to q\bar{q})$ is an irreducible background for this analysis.
In the recoil mass distribution (see the left plot of figure~\ref{fig:2}), the signal and the $\nu\bar{\nu} Z(Z\to q\bar{q})$ backgrounds peak at the mass of the Z boson, and the other SM backgrounds peak at two sides of the distribution.
An optimized cut on the recoil mass has a signal efficiency of 88\% and reduces the background by more than one order of magnitude, see the first row in table~\ref{tab:3}.
The signal has significant invisible energy, so the \emph{visible energy} (visEn, the second row in table~\ref{tab:3}) is about half of the total energy.
The leptonic and semi-leptonic backgrounds have high-energy leptons in the final state, and the fully hadronic backgrounds have a larger multiplicity in the final state.
So, the cut variable of the \emph{leading lepton energy} (leadLepEn, the third row in table~\ref{tab:3}) is used to suppress leptonic and semi-leptonic backgrounds, the cut variable of \emph{multiplicity} (the fourth row in table~\ref{tab:3}) is used to suppress leptonic and some fully hadronic backgrounds.
With the above cuts, 0.01\% of the leptonic backgrounds, 9.24\% of the semi-leptonic backgrounds and 4.82\% of the fully hadronic backgrounds remain in the selected sample.
\begin{figure}[tbp]
\centering 
\includegraphics[width=0.45\textwidth]{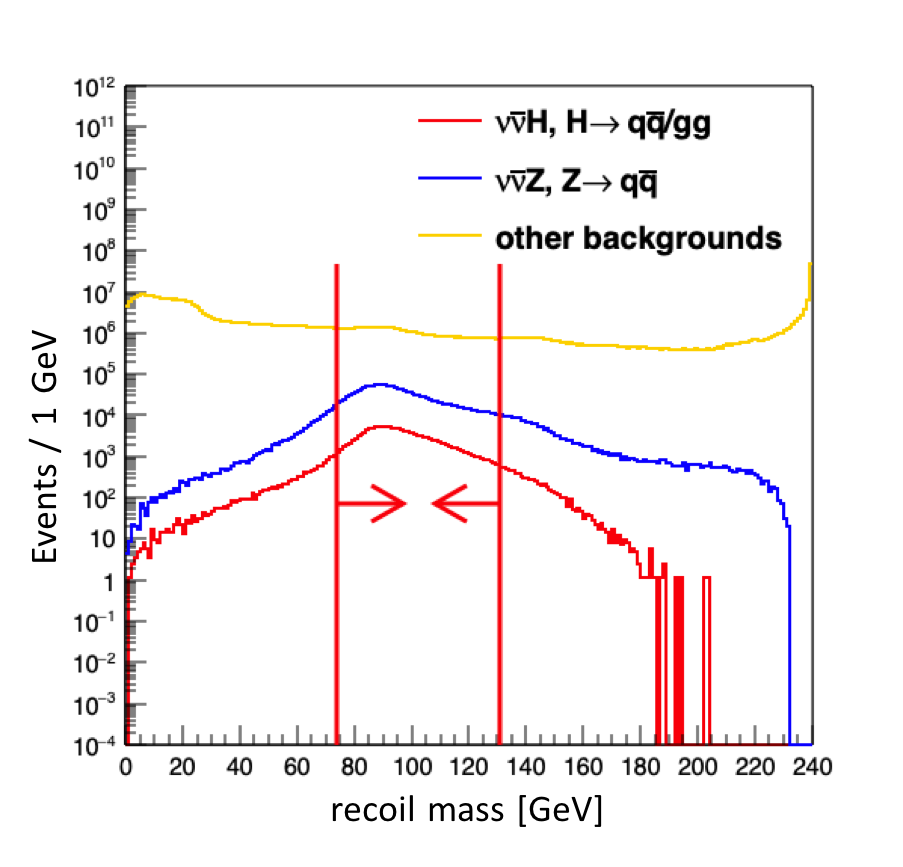}
\hfill
\includegraphics[width=0.45\textwidth]{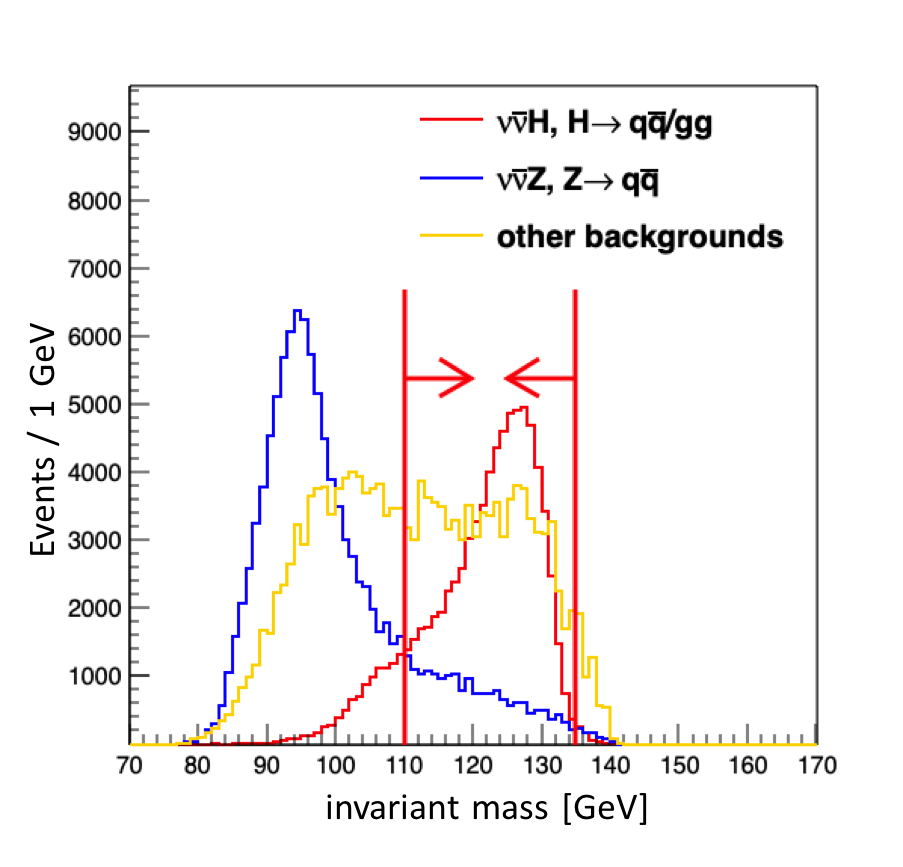}
\caption{\label{fig:2} 
The distributions of the recoil mass of all SM samples and of the invariant mass of SM samples after the Y23 cut for $\nu\bar{\nu} Hq\bar{q}$, $\nu\bar{\nu} Zq\bar{q}$, and other backgrounds are shown in the left plot and right plot, respectively.
}
\end{figure}
\item The remaining backgrounds are dominated by 2f processes consisting of $e^+e^-\to q\bar{q}$.
The $e^+e^-\to q\bar{q}$ backgrounds with high-energy Initial State Radiation (ISR) detected by the detector would have energetic neutral particles in the final state.
Meanwhile, the final state particles of $e^+e^-\to q\bar{q}$ would fly in the end-cap region, so the \emph{visible transverse momentum} ($Pt$, the sixth row in table~\ref{tab:3}) would be lower and the \emph{visible longitudinal momentum} ($Pl$, the seventh row in table~\ref{tab:3}) would be higher than in the signal events.
With the cut variables of the \emph{leading neutral energy} (leadNeuEn, the fifth row in table~\ref{tab:3}), $Pt$, and $Pl$, 0.3\% of the $e^+e^-\to q\bar{q}$ backgrounds remain in the selected sample.
\item A cut on Y23$\footnote{The Durham distance at which a two-jet system can be reconstructed into a three-jet system.}$~\cite{Catani:1991hj, Cacciari:2011ma} is applied to suppress the backgrounds, resulting in an improvement of the signal strength accuracy from 0.94\% to 0.76\%, see the eighth row in table~\ref{tab:3}.
The signal should have an \emph{invariant mass} (InvMass, the ninth row in table~\ref{tab:3}) close to that of the Higgs boson.
After applying the above cut variables, the distributions of invariant mass for signal and backgrounds are shown in the right plot in figure~\ref{fig:2}.
The $\nu\bar{\nu} Z(Z\to q\bar{q})$ backgrounds have their peak at the Z boson, the other SM backgrounds exhibit a flat distribution with statistics comparable to those of $\nu\bar{\nu} Z(Z\to q\bar{q})$ and signal.
The cut variable of the invariant mass could preserve 75\% of the remaining signal events and veto 69\% of the remaining backgrounds.
\end{itemize}

\begin{figure}[tbp]
\centering 
\includegraphics[width=0.45\textwidth]{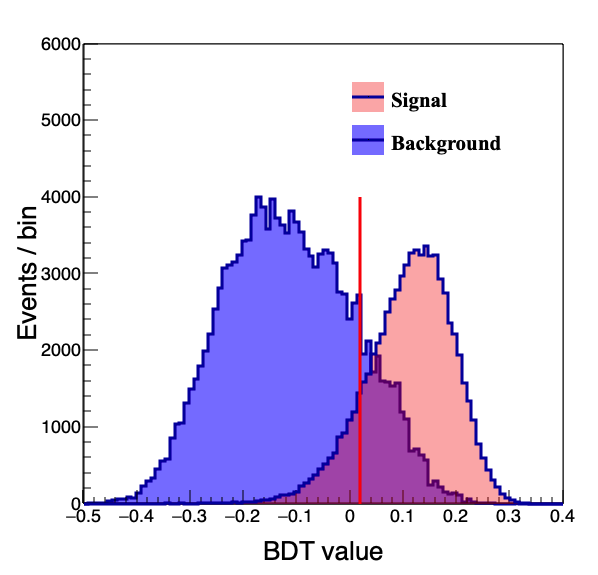}
\caption{\label{fig:BDT} 
BDT output distributions for signal and background events. The samples used here are those passed all the cuts introduced above.
}
\end{figure}

After the cut-based event selection, a Boosted Decision Tree (BDT) is implemented to further improve the selection performance.
The input variables include the cut variables mentioned above and the four-momentum of two jets.
Figure~\ref{fig:BDT} shows the BDT responses. With the optimized cut of the BDT response at 0.02, over 71\% of the backgrounds are rejected at the cost of a 10\% signal loss.
With a signal efficiency of 34\%, the first step reduces background by more than four orders of magnitude, resulting in a relative accuracy of 0.47\% in the measurement of $\nu\bar{\nu} H, H\to q\bar{q}/gg$.
The selection efficiency for $H\to b\bar{b}/c\bar{c}/gg$ is 35\%/33\%/35\%.
The inhomogeneity in $H\to b\bar{b}/c\bar{c}/gg$ selection is mainly caused by the cut variables Y23 and invariant mass, which are discussed in section~\ref{sec:syst}.

The second step focuses on the separation of the different Higgs decay modes and can be divided into two stages.
The first stage aims to obtain the optimized flavor tagging performance matrix (see below) of the CEPC baseline detector, and the  second stage calculate the signal strength accuracy.

\begin{figure}[tbp]
\centering 
\includegraphics[width=0.45\textwidth]{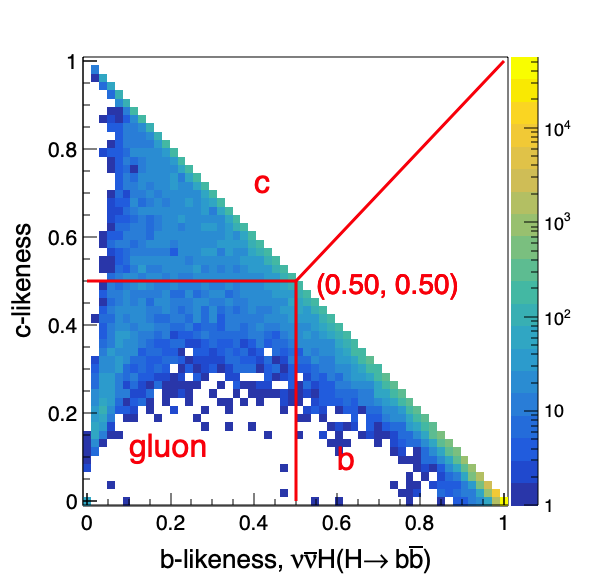}
\hfill
\includegraphics[width=0.45\textwidth]{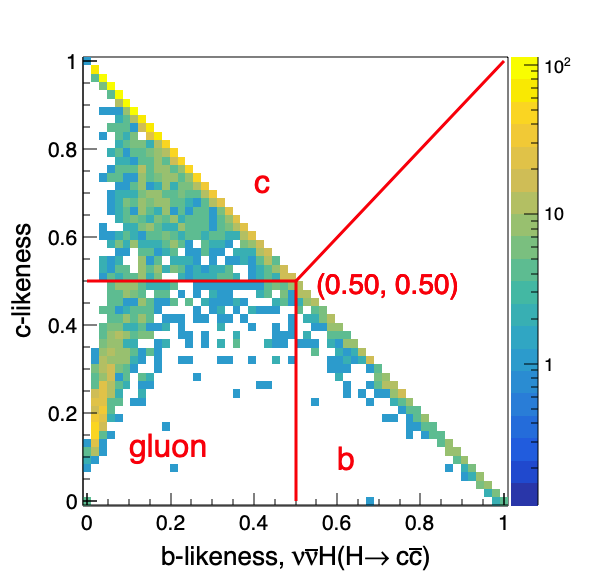}
\\
\includegraphics[width=0.45\textwidth]{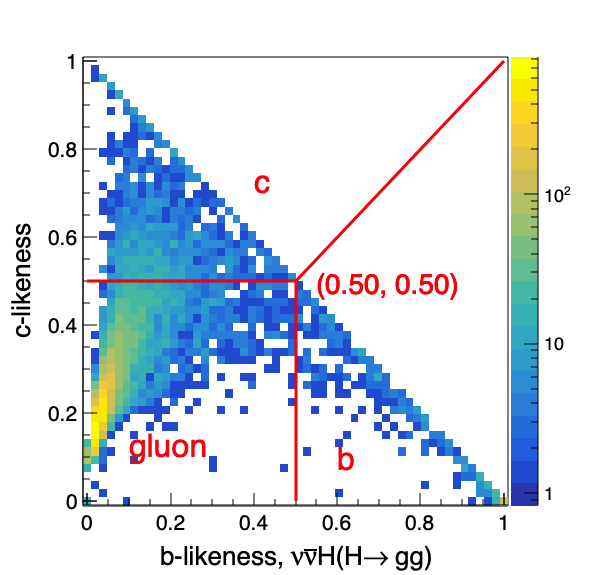}
\hfill
\includegraphics[width=0.45\textwidth]{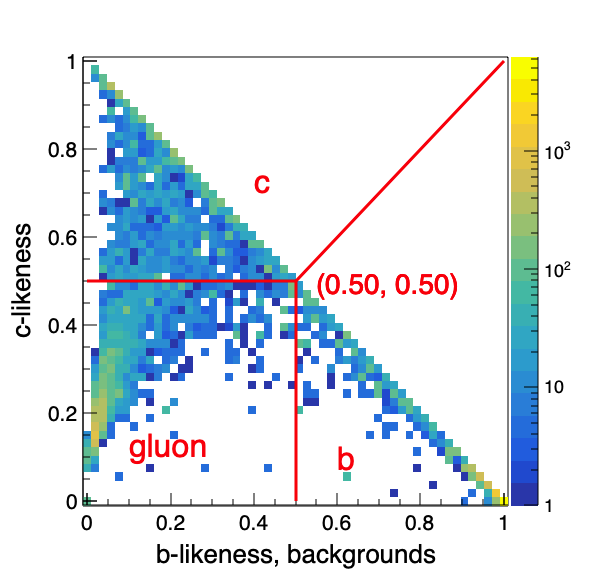}
\\
\includegraphics[width=0.5\textwidth]{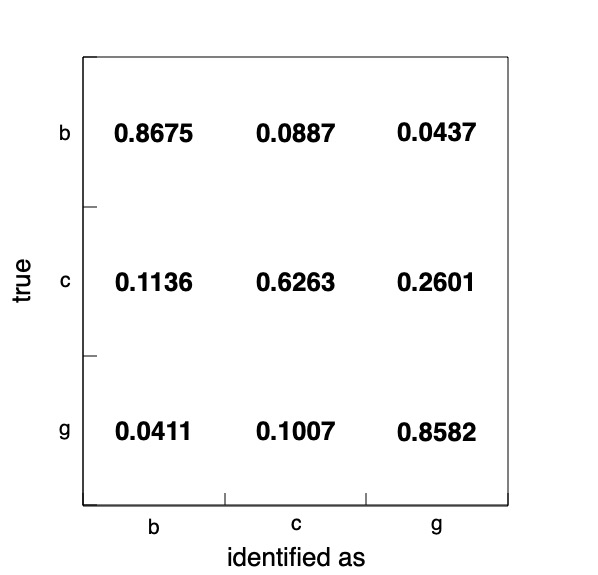}
\caption{\label{fig:3} The distributions of b/c-likeness for $\nu\bar{\nu} H(H\to b\bar{b})$ (top left), $\nu\bar{\nu} H(H\to c\bar{c})$ (top right), $\nu\bar{\nu} H(H\to gg)$ (middle left), and SM backgrounds (middle right). The optimized flavor tagging performance matrix is shown as the bottom plot, where the element represents the flavor identification efficiency.
}
\end{figure}

In the first stage, the particles in the final state are forced into two jets by using the Durham~\cite{Catani:1991hj} jet clustering algorithm implemented in the LCFIPlus software package~\cite{LCFIPlus}. 
The jet clustering algorithm considers single reconstructed particles and composite objects such as reconstructed secondary vertices as basic candidates.
For each jet, the flavor tagging algorithm is used to calculate its likeness to reference samples of b or c jets.
The flavor tagging used in this work is also implemented in the LCFIPlus software package and is performed using Gradient Boosted Decision Tree (GBDT).
The training is applied to the simulated $Z\to q\bar{q}$ sample produced at $\sqrt{s}$ of $91.2\,GeV$.
The reconstructed jets in the sample are divided into 4 categories depending on the number of secondary vertices and isolated leptons in the jet: jets with secondary vertex and lepton, jets with secondary vertex but without lepton, jets without secondary vertex but with lepton and jets without secondary vertex and lepton.
In each category, two types of flavor tagging algorithms are trained using the GBDT method, one for the b-tagging algorithm and the other for the c-tagging algorithm.

\begin{figure}[tbp]
\centering 
\includegraphics[width=0.45\textwidth]{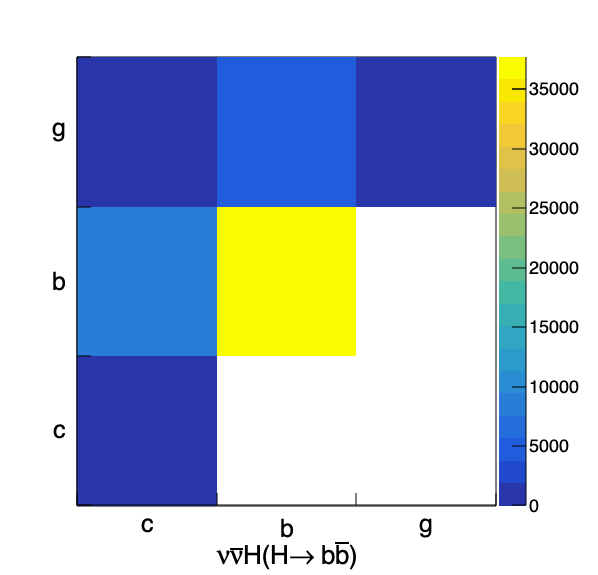}
\hfill
\includegraphics[width=0.45\textwidth]{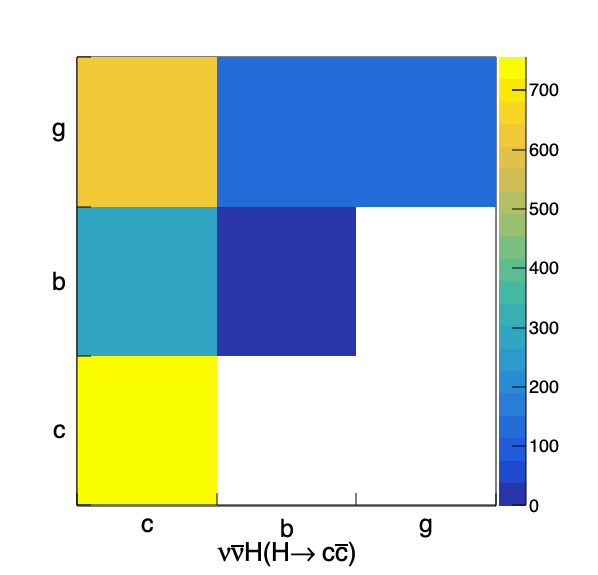}
\\
\includegraphics[width=0.45\textwidth]{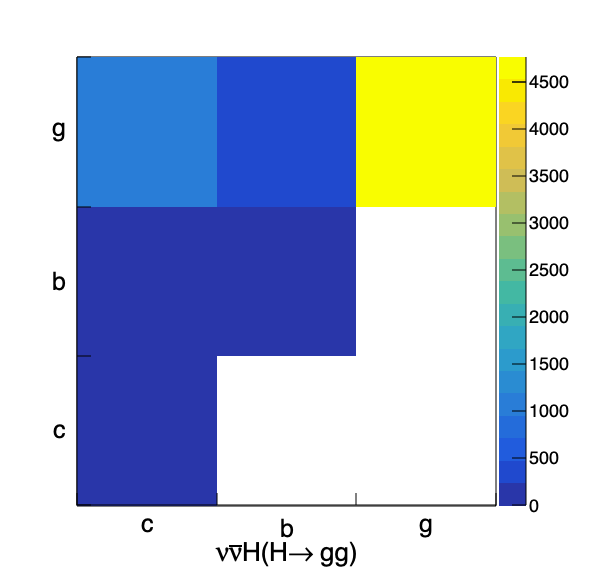}
\hfill
\includegraphics[width=0.45\textwidth]{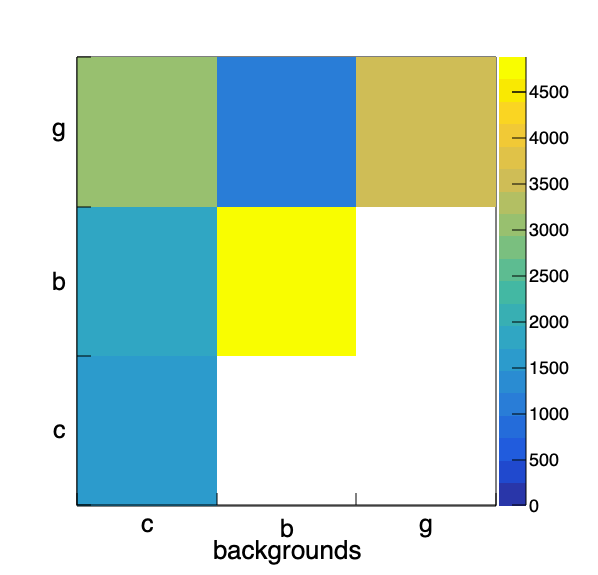}
\caption{\label{fig:5}  The distributions of $\nu\bar{\nu} H(H\to b\bar{b})$ (top left), $\nu\bar{\nu} H(H\to c\bar{c})$ (top right), $\nu\bar{\nu} H(H\to gg)$ (bottom left), and backgrounds (bottom right) based on the optimized flavor tagging performance matrix.}
\end{figure}

The distributions of b/c-likeness are shown in figure~\ref{fig:3} for $\nu\bar{\nu} H(H\to b\bar{b})$, $\nu\bar{\nu} H(H\to c\bar{c})$, $\nu\bar{\nu} H(H\to gg)$, and the remaining SM backgrounds.
The phase space spanned by the b/c-likeness is divided into three different regions corresponding to the identified b, c, and gluons.
We then obtain the ratio of b-jet identified as b-jet, b-jet identified as c-jet, and so on. 
These ratios can be represented with a migration matrix, the form of which is shown in figure~\ref{fig:3}.
We optimize the working point (phase space separation) to maximize the trace of the migration matrix. 
The optimized migration matrix is shown as the bottom plot in figure~\ref{fig:3}.
In principle, the working point can be optimized independently for $H\to b\bar{b}$, $c\bar{c}$, and $gg$ measurements.
We evaluate the corresponding performance and find that the final accuracy can be improved by sub-percent level.
Since the improvement is not significant, a uniform matrix for $\nu\bar{\nu} H(H\to b\bar{b}/c\bar{c}/gg)$ is used for simplicity.
According to the identified jet-flavor combinations, the signal events and backgrounds are classified into six different categories (see figure~\ref{fig:5}).

In the second stage, the relative accuracy of the signal strength could be calculated by the log-likelihood function~\eqref{eq:loglikelihood}~\cite{Zyla:2020zbs, MLE}, 
\begin{equation}
\label{eq:loglikelihood}
-2\cdot log(\ell) = \sum_{i=1}^{i = 6}\frac{ [ S_{b} \cdot N_{b, i} + S_{c} \cdot N_{c, i} + S_{g} \cdot N_{g, i} + N_{bkg, i} - N_{i}    ]^2  }{N_i},
\end{equation}
where $S_{b}$ represents the signal strength of $\nu\bar{\nu} H(H\to b\bar{b})$, 
$N_{b, i}$ represents the event count of $\nu\bar{\nu} H(H\to b\bar{b})$ in the $ith$ bin, 
$N_{bkg, i}$ represents the event count of the backgrounds in the $ith$ bin,
and $N_{i}$ represents the total event count ($\nu\bar{\nu} H$ with $H\to b\bar{b}/c\bar{c}/gg$ and backgrounds) in the $ith$ bin,
similar for $S_{c}$, $S_{g}$, $N_{c, i}$, and $N_{g, i}$.
The error covariance matrix is obtained from the Hessian matrix of the log-likelihood function with respect to three signal strengths. 
The relative accuracies of the signal strengths are the square roots of the diagonal elements of the covariance matrix.
It is 0.49\%/5.75\%/1.82\% for $\nu\bar{\nu} H(H\to b\bar{b}/c\bar{c}/gg)$.

\subsection{ \texorpdfstring{$q\bar{q}H$}.}
\label{sec:qqH}
In this subsection, the accuracy of the $q\bar{q}H(H\to b\bar{b}/c\bar{c}/gg)$ signal strength is analyzed.
The analysis process is similar to that in the $\nu\bar{\nu} H$ channel.
Since the backgrounds consist of leptonic, semi-leptonic, and fully hadronic samples,
the first step can be divided into three stages to select the signal events step by step.

\begin{enumerate}
\item Selection of fully hadronic events.
\item Selecting events with 4-jet topology.
\item Selection of ZH events.
\end{enumerate}

\begin{table}[tbp]
\centering
\begin{scriptsize}
\begin{tabular}{|ccccccccccc|}
\hline
                                          & $q\bar{q}Hq\bar{q}/gg$       &2f                   & SW              &SZ               &WW             &ZZ              &Mixed          &ZH    & $\gamma\gamma$     &$\frac{\sqrt{S+B}}{S}(\%)$      \\
total                                   &  527488     &$8.01E8$      &$1.95E7$    &$9.07E6$    & $5.08E7$    &$6.39E6$   &$2.18E7$      &$613008$        &     $4.91E8$     & 7.10                                       \\
\hline
multiplicity                         & \multirow{2}{*}{527488}      &\multirow{2}{*}{$3.04E8$}     &\multirow{2}{*}{$1.46E7$}     &\multirow{2}{*}{$3.37E6$}     &\multirow{2}{*}{$4.85E7$}    &\multirow{2}{*}{$6.00E6$}                  &\multirow{2}{*}{$1.81E7$}     &\multirow{2}{*}{$577930$}       &  \multirow{2}{*}{$4.12E8$}     & \multirow{2}{*}{5.39}    \\
$\in(27, +\infty)$                &&&&&&&&&& \\                         
$leadLepEn$  (GeV)                        &\multirow{2}{*}{527036}       &\multirow{2}{*}{$2,98E8$}      &\multirow{2}{*}{$6.76E6$}    &\multirow{2}{*}{$2.44E6$}     &\multirow{2}{*}{$3.93E7$}    &\multirow{2}{*}{$5.40E6$}      &\multirow{2}{*}{$1.79E7$}     &\multirow{2}{*}{$531411$}     &   \multirow{2}{*}{$4.12E8$}    & \multirow{2}{*}{5.31}    \\
$\in(0, 59)$                            &&&&&&&&&& \\ 
$visEn$  (GeV)                            & \multirow{2}{*}{510731}     &\multirow{2}{*}{$1.21E8$}      &\multirow{2}{*}{$1.29E6$}     &\multirow{2}{*}{$551105$}     &\multirow{2}{*}{$2.14E7$}    &\multirow{2}{*}{$3.06E6$}     &\multirow{2}{*}{$1.71E7$}     &\multirow{2}{*}{$180571$}        &  \multirow{2}{*}{22643}   & \multirow{2}{*}{2.52}    \\
$\in(199, 278)$                            &&&&&&&&&& \\ 
\hline
$leadNeuEn$  (GeV)                       &\multirow{2}{*}{509623}       &\multirow{2}{*}{$5.68E7$}      &\multirow{2}{*}{$716161$}    &\multirow{2}{*}{$168030$}    &\multirow{2}{*}{$2.04E7$}     &\multirow{2}{*}{$2.93E6$}     &\multirow{2}{*}{$1.65E7$}     &\multirow{2}{*}{$176387$}     &  \multirow{2}{*}{21205}   &\multirow{2}{*}{1.94}     \\
$\in(0, 57)$                            &&&&&&&&&& \\ 
$thrust$                            & \multirow{2}{*}{460535}      &\multirow{2}{*}{$7.81E6$}      &\multirow{2}{*}{$473732$}    &\multirow{2}{*}{$132126$}    &\multirow{2}{*}{$1.88E7$}     &\multirow{2}{*}{$2.60E6$}     &\multirow{2}{*}{$1.54E7$}     &\multirow{2}{*}{$167863$}     &   \multirow{2}{*}{6110}   &\multirow{2}{*}{1.47}      \\
$\in(0, 0.86)$                            &&&&&&&&&& \\ 
$-log(Y_{34})$                 & \multirow{2}{*}{451468}      &\multirow{2}{*}{$4.90E6$}       &\multirow{2}{*}{$181432$}    &\multirow{2}{*}{$119836$}     &\multirow{2}{*}{$1.74E7$}    &\multirow{2}{*}{$2.40E6$}      &\multirow{2}{*}{$1.45E7$}    &\multirow{2}{*}{$165961$}      &  \multirow{2}{*}{4672}    &\multirow{2}{*}{1.40}    \\
$\in(0, 5.8875)$                            &&&&&&&&&& \\ 
\hline
$HiggsJetsA$                              &  \multirow{2}{*}{326207}       & \multirow{2}{*}{$2.83E6$}      & \multirow{2}{*}{$110156$}     & \multirow{2}{*}{58613}         & \multirow{2}{*}{$4.54E6$}      &  \multirow{2}{*}{$870276$}   &  \multirow{2}{*}{$3.74E6$}    &  \multirow{2}{*}{96560}     &   \multirow{2}{*}{2156}    &  \multirow{2}{*}{1.08}        \\
$\in(2.18, 4)$                            &&&&&&&&&& \\ 
$ZJetsA$                              & \multirow{2}{*}{279030}        & \multirow{2}{*}{$1.37E6$}      & \multirow{2}{*}{$33491$}       & \multirow{2}{*}{37101}         & \multirow{2}{*}{$2.39E6$}      & \multirow{2}{*}{496611}         & \multirow{2}{*}{$2.00E6$}     & \multirow{2}{*}{74005}        &    \multirow{2}{*}{1797}     &  \multirow{2}{*}{0.93}    \\
$\in(1.97, 4)$                            &&&&&&&&&& \\ 
$ZHiggsA$                                & \multirow{2}{*}{274530}        & \multirow{2}{*}{$1.32E6$}      & \multirow{2}{*}{$17026$}       & \multirow{2}{*}{33847}         & \multirow{2}{*}{$2.28E6$}      & \multirow{2}{*}{468340}         & \multirow{2}{*}{$1.91E6$}     & \multirow{2}{*}{69620}       &    \multirow{2}{*}{1797}    &  \multirow{2}{*}{0.92}    \\
$\in(2.32, 4)$                            &&&&&&&&&& \\ 
$circle$                           &268271       & 1.20E6         & 10193           & 31567        & $2.13E6$     & 424514        & 1.79E6       & 65434     &   $0^*$     & 0.90             \\
\hline
BDT                                & \multirow{2}{*}{192278}       &  \multirow{2}{*}{378300}         &  \multirow{2}{*}{40}                &  \multirow{2}{*}{307}             &  \multirow{2}{*}{271436}        &  \multirow{2}{*}{141446}        &  \multirow{2}{*}{244126}       &  \multirow{2}{*}{30022}         &    \multirow{2}{*}{$0^*$}     &  \multirow{2}{*}{0.57}             \\
$\in(0.02, 1)$ &&&&&&&&&& \\ 
\hline
\end{tabular}
\end{scriptsize}
\caption{\label{tab:4} The event selection of $q\bar{q} H(H\to q\bar{q}/gg)$ is based on the integrated luminosity of $5.6\,ab^{-1}$. The $\gamma\gamma$ is the abbreviation for $\gamma\gamma\to hadrons$ process. The symbol $0^*$ represents that the number of $\gamma\gamma\to hadrons$ events is less than $3.09/0.0028$ at a confidence level of 95\% according to Feldman-Cousins approach in the case of observing a zero event, where 3.09 is quoted from ref.~\cite{Feldman:1997qc} and 0.0028 is the scaling factor of the $\gamma\gamma\to hadrons$ process.}
\end{table}

The cutflow corresponding to these three stages is given in table~\ref{tab:4}.
The $\gamma\gamma$ is the abbreviation of $\gamma\gamma\to hadrons$ process.
The first stage aims to suppress the leptonic and semi-leptonic backgrounds that have low multiplicity or high-energy leptons ($e^\pm/\mu^\pm$) or invisible leptons ($\nu/\bar{\nu}$).
Thus, with the cut variables of \emph{multiplicity} (the first row in table~\ref{tab:4}), the \emph{leading lepton energy} (leadLepEn, the second row in table~\ref{tab:4}), and the \emph{visible energy} (visEn, the third row in table~\ref{tab:4}), the background statistic is reduced to 26\%.
The cut variable of visible energy can also veto some $e^+e^-\to q\bar{q}$ backgrounds with high-energy ISR that have escaped the detector.
Step to the second stage: 
the cut variables of the \emph{leading neutral energy} (leadNeuEn, the fourth row in table~\ref{tab:4}, aims to suppress $e^+e^-\to q\bar{q}$ with high-energy ISR detected by the detector), \emph{thrust}$\footnote{To evaluate the thrust of an event, first determine the thrust axis $n_T$, which is the direction of maximum momentum flow. The thrust is then defined as the fraction of the particle momentum that flows along the thrust axis. }$~\cite{ATLAS:2020vup} (the fifth row in table~\ref{tab:4}), and Y34 $\footnote{The Durham distance at which a three-jet system can be reconstructed into a four-jet system.}$ (the sixth row in table~\ref{tab:4}), are used to select 4-quark samples from the full hadronic samples.
The second stage reduces the remaining background by almost an order of magnitude at the cost of losing 11\% of the signal events.

\begin{figure}[tbp]
\centering 
\includegraphics[width=.45\textwidth]{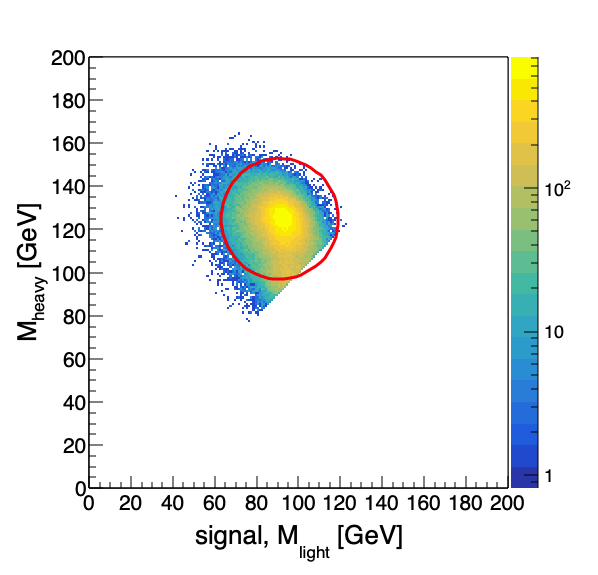}
\hfill
\includegraphics[width=.45\textwidth]{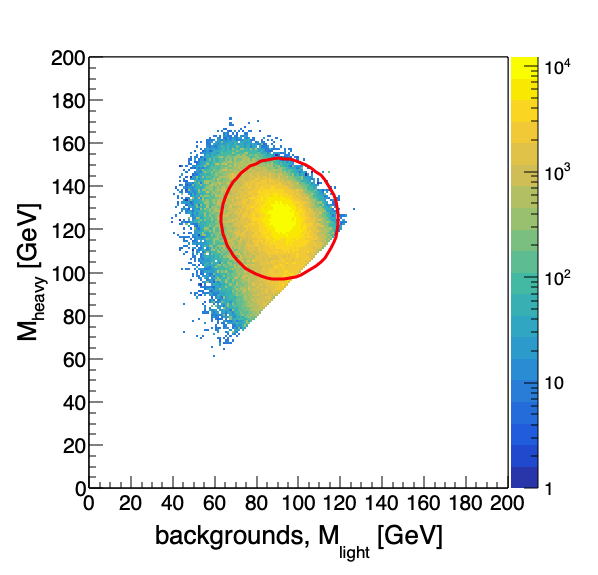}
\caption{\label{fig:6} The mass distribution of two di-jet systems, the left plot refers to the signal, the right to the backgrounds.}
\end{figure}

Since the signal contains only four jets, the particles in the final state are first forced into four jets using Durham algorithm.
There are two bosons in the signal, then the four jets in the final state are paired into two di-jet systems using the pairing method of the minimization eq.~\eqref{eq:pair},
\begin{equation}
\label{eq:pair}
\chi^2 = \frac{(M_{12} - M_{B1})^2}{\sigma_{B1}^2} + \frac{(M_{34} - M_{B2})^2}{\sigma_{B2}^2},
\end{equation}
where $M_{12}$ and $M_{34}$ are the masses of the di-jet systems and $M_{B1}$ and $M_{B2}$ are the reference masses of the Z or W or the Higgs boson.
The $\sigma$ is the convolution of the boson width and the detector resolution.
According to~\cite{CEPCStudyGroup:2018ghi}, the detector resolution is 4\% of the boson mass.
After pairing four jets into two di-jet systems, we refer to the di-jet system with heavy invariant mass ($M_{heavy}$) as the heavy di-jet system and the one with light invariant mass ($M_{light}$) as the light di-jet system.
There are angular variables that can be used to separate the signal from the remaining backgrounds. 
Angular variables include: the angle between two jets of a light di-jet system (ZJetsA, the seventh row in table~\ref{tab:4}), the angle between two jets of a heavy di-jet system (HiggsJetsA, the eighth row in table~\ref{tab:4}), and the angle between two di-jet systems (ZHiggsA, the ninth row in table~\ref{tab:4}).
These three cut variables could reduce more than 84\% (from $3.97\times 10^7$ to $6.10\times 10^6$) of the backgrounds.
For the signal, one di-jet system should have an invariant mass near the Higgs boson and the other should have an invariant mass near the Z boson.
Then, a circular selection (the tenth row in table~\ref{tab:4}) $(M_{heavy} - 125)^2 + (M_{light} - 91)^2 <= 29^2$ can be used to select signal events, as shown in figure~\ref{fig:6}.

\begin{figure}[tbp]  
\centering 
\includegraphics[width=0.5\textwidth]{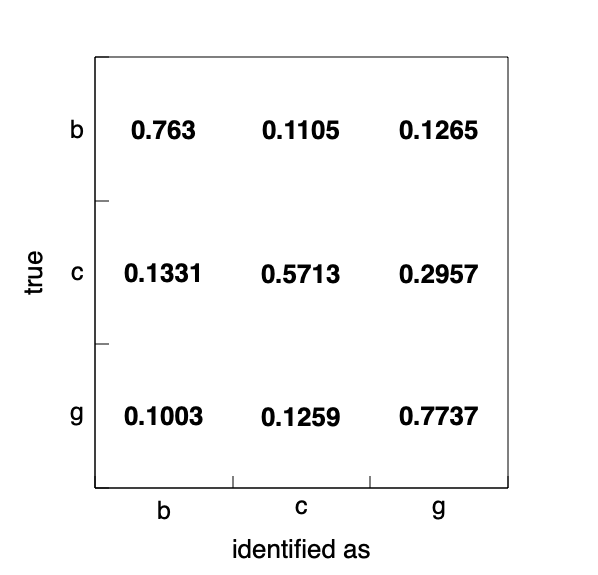}
\\
\includegraphics[width=0.45\textwidth]{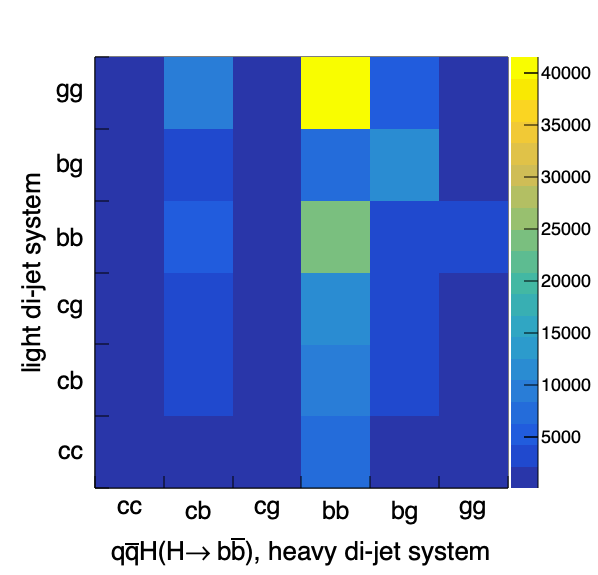}
\hfill
\includegraphics[width=0.45\textwidth]{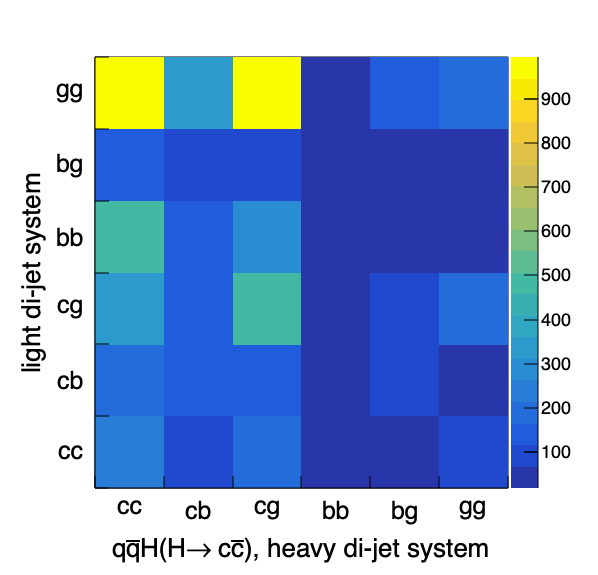}
\\
\includegraphics[width=0.45\textwidth]{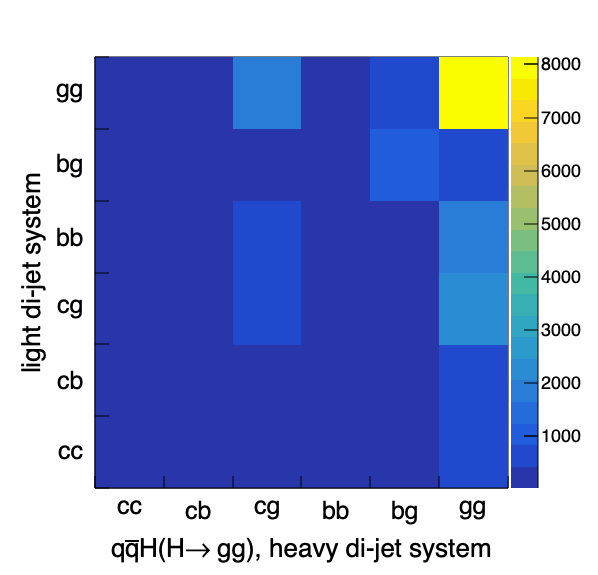}
\hfill
\includegraphics[width=0.46\textwidth]{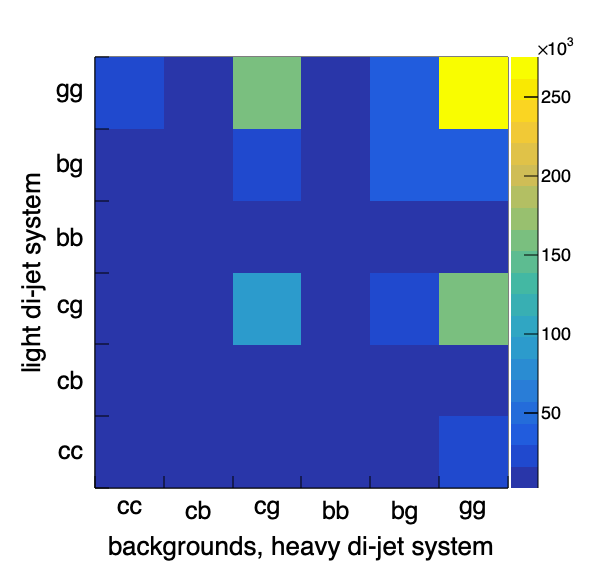}
\caption{\label{fig:qqHcate} The optimized flavor tagging performance matrix for heavy di-jet system (upper-left) and light di-jet system (upper-right).  The distributions of $q\bar{q} H(H\to b\bar{b})$ (middle left), $q\bar{q} H(H\to c\bar{c})$ (middle right), $q\bar{q} H(H\to gg)$ (bottom left), and backgrounds (bottom right).}
\end{figure}

To fully exploit the characteristics of the signal and backgrounds, a BDT method is used to suppress the backgrounds. 
Input variables include the cut variables mentioned above, the four-momentum of four jets, and several event shape variables \cite{ATLAS:2020vup} including max-broadening$\footnote{The max-broadening is related to the transverse momentum measured with respect to the thrust axis. Introduce a plane perpendicular to the thrust axis and divide the space into two hemispheres. The jet broadening of each hemisphere is defined as $B = \frac{1}{2\sum_{j=1}^{N_{particles}}|P_j|}\sum_{i=1, P_{i}\cdot n_{T}>0}^{N_{particles}}|P_{i}\times n_{T}|$, where $P_{i}$ is the 3-momentum of particle i, $n_T$ is the thrust axis, and $P_{i}\cdot n_{T}>0(P_{i}\cdot n_{T}<0)$ is used to divide the particles into two hemispheres. The max-broadening is then the maximum jet broadening between these two hemispheres.}$, C-parameter$\footnote{The linearized sphericity is defined as $L^{ab} = \frac{1}{\sum_{j=1}^{N_{particles}}|P_{j}|}\sum_{i=1}^{N_{particles}}\frac{P_{i}^aP_{i}^b}{|P_{i}|}$, where $P_{i}$ is the 3-momentum of particle i, and $P_{i}^a$ denotes the component a of the 3-momentum of the particle i. Then the C-parameter can be calculated as $C = 3(\lambda_{1}\lambda_{2} + \lambda_{1}\lambda_{3} + \lambda_{2}\lambda_{3})$, where  $\lambda$ is the eigenvalue of $L^{ab}$.}$, and D-parameter$\footnote{The D-parameter can be calculated as $D = 27\cdot \lambda_1\cdot \lambda_2\cdot \lambda_3$. }$.
Finally,  the total SM background is reduced to 1.07 million statistics, and more than 36\% of the total $q\bar{q}H(H\to q\bar{q}/gg)$ signal events survived, resulting in a relative uncertainty of 0.57\%.

After the event selection process, an optimized flavor tagging performance matrix can be found by setting an optimized working point on the distributions of b/c-likeness of two jets from the heavy di-jet system, which is shown in the top plot of figure~\ref{fig:qqHcate}.
Compared to $\nu\bar{\nu} H$, the diagonal elements have decreased by 13\%/8\%/10\% for $b/c/g$.
In other words, the identification performance of the $b/c/g$ jet in the $q\bar{q}H$ channel is slightly worse than that in the $\nu\bar{\nu} H$ channel.
This is due to the visible particles decaying from the Z boson would degrade the jet clustering performance.
Based on the optimized flavor tagging performance matrix, the identified flavor combinations of $q\bar{q}H(H\to b\bar{b})$, $q\bar{q}H(H\to c\bar{c})$, $q\bar{q}H(H\to gg)$, and backgrounds are shown in figure~\ref{fig:qqHcate}.
The x and y axes represent the flavor of two jets from the heavy and light di-jet systems, respectively.
The light di-jet system in the signal events corresponds to the Z-boson.
Due to the imperfect flavor tagging performance, the pattern of Z-boson decay in figure~\ref{fig:qqHcate} is not clear.
Using the log-likelihood function similar to that of $\nu\bar{\nu} H$, the relative accuracy for $q\bar{q}H(H\to b\bar{b}/c\bar{c}/gg)$ is calculated to be 0.35\%/7.74\%/3.96\%. 

\subsection{Combination}
To first order, the relative accuracy of $H\to b\bar{b}/c\bar{c}/gg$ signal strength is measured independently in three channels, $\nu\bar{\nu} H$, $q\bar{q}H$, and $\ell^+\ell^-H$.
At a centre-of-mass energy of $240\,GeV$ and an integrated luminosity of 5.6 $ab^{-1}$, the relative statistical accuracy of $H\to b\bar{b}/c\bar{c}/gg$ signal strength can reach 0.27\%/4.03\%/1.56\% when combined with these three channels.
The results of the analysis are summarized in table~\ref{tab:5}.
According to the recently released Snowmass~\cite{Gao:2022lew}, the CEPC operates at $240\,$GeV will integrate $20\,ab^{-1}$ luminosity.
Accordingly, the relative statistical accuracy of the $H\to b\bar{b}/c\bar{c}/gg$ signal strength would be 0.14\%/2.13\%/0.82\%.

 \begin{table}[tbp]
\centering
\begin{tabular}{|cccc|}
\hline
Z decay mode          &   $H\to b\bar{b}$      & $H\to c\bar{c}$     & $H\to gg$    \\
$Z\to e^+e^-$           &   1.57\%                    & 14.43\%                  & 10.31\%    \\
$Z\to \mu^+\mu^-$   &  1.06\%                     & 10.16\%                  & 5.23\%     \\
$Z\to q\bar{q}$         & 0.35\%                    & 7.74\%                  & 3.96\%   \\
$Z\to \nu\bar{\nu}$   & 0.49\%                    & 5.75\%                  & 1.82\%   \\
combination             &  0.27\%                    &  4.03\%                            & 1.56\%          \\

\hline
\end{tabular}
\caption{\label{tab:5} The signal strength accuracies for different channels.}
\end{table}

\section{Dependence of accuracies on critical detector performances}
\label{sec:Opt}

The flavor tagging performance and the color-singlet-identification (CSI), which represents the reconstruction of a color-singlet that decays into two jets, are two critical detector performances for measuring the signal strength accuracy of $H\to b\bar{b}/c\bar{c}/gg$.
In the $\nu\bar{\nu} H$ channel, the impact of the flavor tagging performance is analyzed as shown in subsection~\ref{sec:OptvvH}.
In the $q\bar{q}H$ channel, there are four jets from the two bosons, so the critical detector performance includes not only flavor tagging but also the CSI. 
Their impact on the anticipated physics reach is evaluated in subsection~\ref{sec:OptqqH}.

\subsection{ \texorpdfstring{$\nu\bar{\nu} H$}.: Flavor tagging}
\label{sec:OptvvH}

The flavor tagging performance can be described by the migration matrix, which is defined by the bottom plot in figure~\ref{fig:3}.
We have three reference points for the migration matrix:
the unitary matrix corresponding to perfect flavor tagging performance, 
the flat matrix (all elements are equal to one-third) corresponding to the performance without flavor tagging,
and the matrix corresponding to the CEPC baseline detector, which is shown as the bottom plot in figure~\ref{fig:3}.

\begin{figure}[tbp]
\centering 
\includegraphics[width=.45\textwidth]{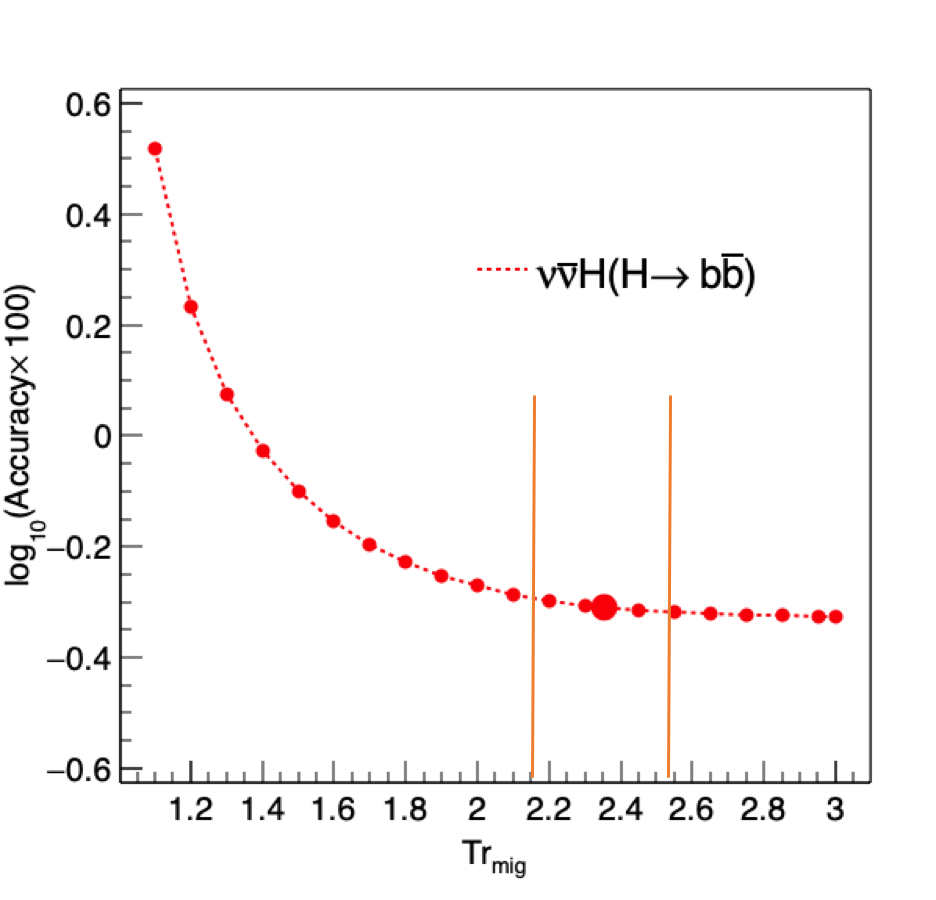}
\hfill
\includegraphics[width=.45\textwidth]{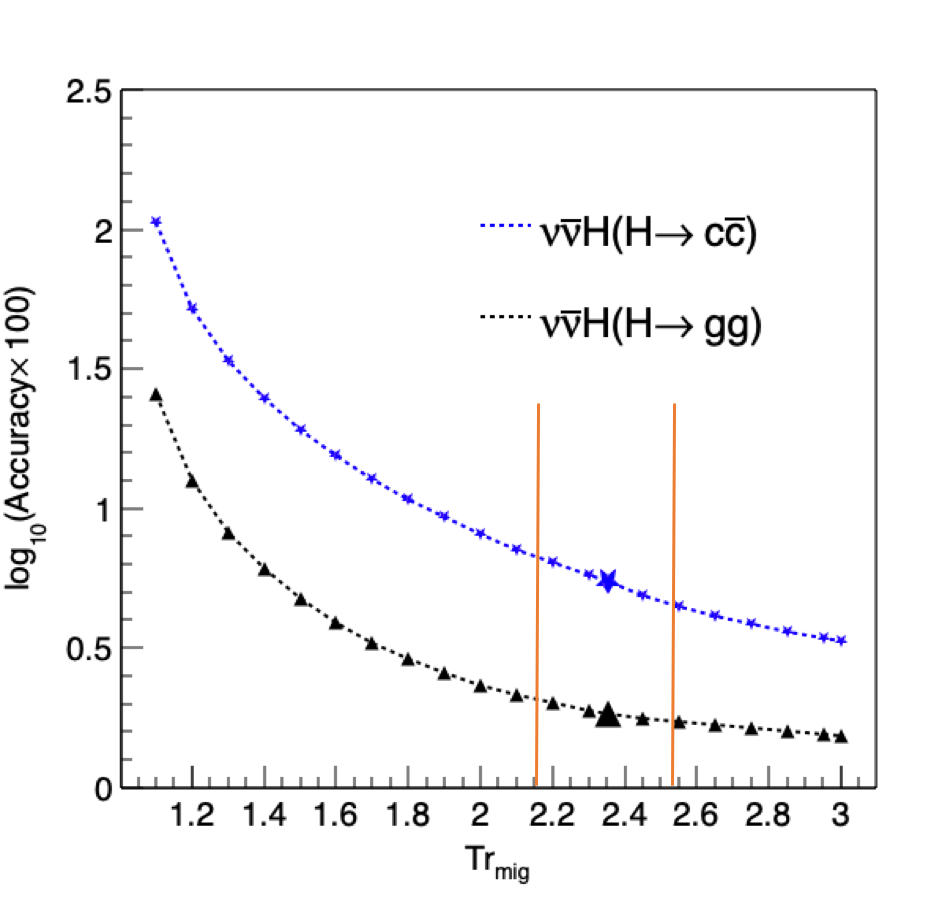}
\caption{\label{fig:11} The dependence of $\nu\bar{\nu} H(H\to b\bar{b}/c\bar{c}/gg)$ signal strength accuracy on flavor tagging performance is shown in this figure, the left plot is for $\nu\bar{\nu} H(H\to b\bar{b})$, and the right plot is for $\nu\bar{\nu} H(H\to c\bar{c}/gg)$. The bigger markers correspond to the results of the CEPC baseline detector. 
When the key vertex detector parameters, including inner radius,  material budget, and spatial resolution, are changed by a factor 0.5/2 from the baseline design (the geometry we used in this simulation), the $Tr_{mig}$ value changes accordingly from 2.35 to 2.54/2.16, shown as two vertical orange lines.}
\end{figure}

An interpolation method is used to obtain different flavor tagging performance matrices, shown as the eq.~\eqref{eq:FL},
\begin{equation}
\label{eq:FL}
\begin{split}
M_{mig} &= \frac{Tr_{mig} - Tr_{opt}}{Tr_{I} - Tr_{opt}}\cdot (M_{I} - M_{opt}) + M_{opt}  \\
M_{mig} &= \frac{Tr_{mig} - Tr_{opt}}{Tr_{1/3} - Tr_{opt}}\cdot (M_{1/3} - M_{opt}) + M_{opt}
\end{split}
\end{equation}
where $M_{I}$ represents the perfect flavor tagging performance matrix (identity matrix), 
$M_{1/3}$ represents the matrix without flavor tagging (all elements equal $1/3$),
$M_{opt}$ represents the flavor tagging performance of the CEPC baseline detector,
$Tr_{I}$ represents the trace of the perfect flavor tagging performance matrix, 
and $Tr_{1/3}$ represents the trace of the matrix without flavor tagging.
$Tr_{mig}$ is a variable whose value and the temporary matrix ($M_{mig}$) have a one-to-one relationship.
The value of $Tr_{opt}$ ranges from $Tr_{1/3}$ to $Tr_{I}$.
If $Tr_{mig}$ is greater than $Tr_{opt}$, use the upper formula of eq.~\eqref{eq:FL}.
Else we use the lower formula.
The value of $Tr_{mig}$ is varied from 1.0 (without flavor tagging) to 3.0 (perfect flavor tagging) in increments of 0.1.
The dependence of signal strength accuracy on the flavor tagging performance is shown in figure~\ref{fig:11}.
Accuracies corresponding to the baseline CEPC detector are represented by the markers at $Tr_{mig}$ = 2.34.
With an ideal flavor tagging algorithm, the signal strength accuracy is 0.48\%/3.53\%/1.61\% for $\nu\bar{\nu} H(H\to b\bar{b}/c\bar{c}/gg)$, which is a 2\%/63\%/13\% improvement over the baseline CEPC detector (0.49\%/5.75\%/1.82\%).
The performance of the flavor tagging depends on the design of the vertex detector. 
If the key vertex detector parameters, including the inner radius,  material budget, and spatial resolution, are changed by a factor 0.5/2~\cite{VTX} from the baseline design (the geometry we used in this simulation), the $Tr_{mig}$ value changes accordingly from 2.35 to 2.54/2.16, as shown in figure~\ref{fig:11}.

\begin{figure}[tbp]
\centering 
\includegraphics[width=.45\textwidth]{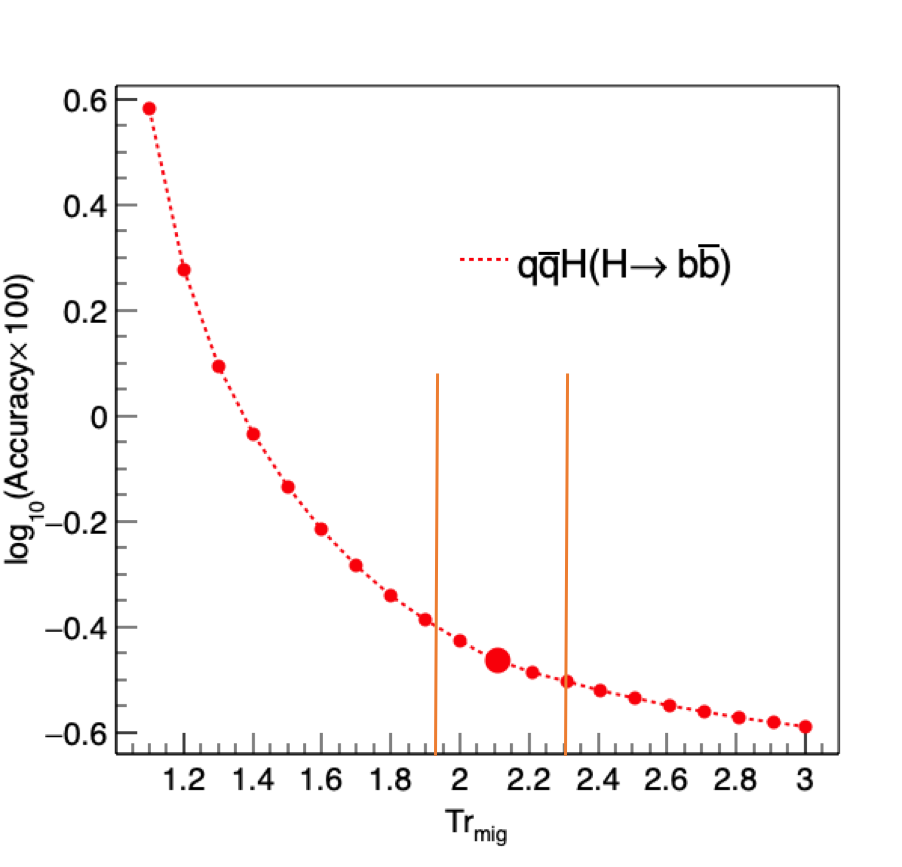}
\hfill
\includegraphics[width=.45\textwidth]{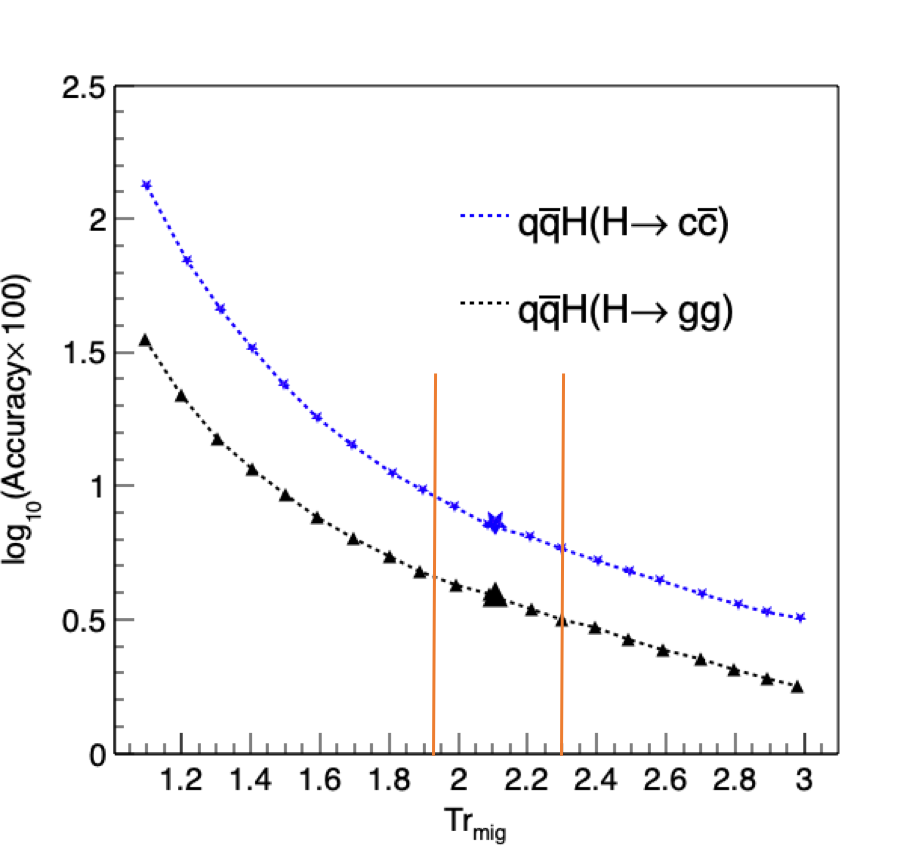}
\caption{\label{fig:12} The dependence of $q\bar{q} H(H\to b\bar{b}/c\bar{c}/gg)$ signal strength accuracy on flavor tagging performance is shown in this figure, the left plot is for $q\bar{q} H(H\to b\bar{b})$, and the right plot is for $q\bar{q} H(H\to c\bar{c}/gg)$. The bigger markers correspond to the results of the CEPC baseline detector.
When the key vertex detector parameters, including inner radius,  material budget, and spatial resolution, are changed by a factor 0.5/2 from the baseline design, the $Tr_{mig}$ value changes accordingly from 2.12 to 2.31/1.93, shown as two vertical orange lines.}
\end{figure}

\subsection{ \texorpdfstring{$q\bar{q}H$}.: Flavor tagging \& CSI}
\label{sec:OptqqH}
Similar to $\nu\bar{\nu} H$, the dependence of the $q\bar{q}H(H\to b\bar{b}/c\bar{c}/gg)$ signal strength accuracy on the flavor tagging performance is shown in figure~\ref{fig:12}.
With perfect flavor tagging performance, the relative accuracy is 0.26\%/3.48\%/1.41\% for $q\bar{q}H (H\to b\bar{b}/c\bar{c}/gg)$, which is a 35\%/122\%/181\% improvement over the baseline CEPC detector (0.35\%/7.74\%/3.96\%).
There is a significant improvement for $q\bar{q}H (H\to gg)$, because after event selection the backgrounds consist mainly of the processes of $e^+e^-\to q\bar{q}/W^+W^-/ZZ/Mixed$ with c-jets/light-jets in the final state, while the CEPC baseline performance classifies almost all light-jets and 30\% of c-jets as gluon-jets.
When the key vertex detector parameters, including inner radius,  material budget, and spatial resolution, are changed by a factor 0.5/2~\cite{VTX} from the baseline design, the $Tr_{mig}$ value changes accordingly from 2.12 to 2.31/1.93, as shown in figure~\ref{fig:12}.

\begin{figure}[tbp]
\centering 
\includegraphics[width=.45\textwidth]{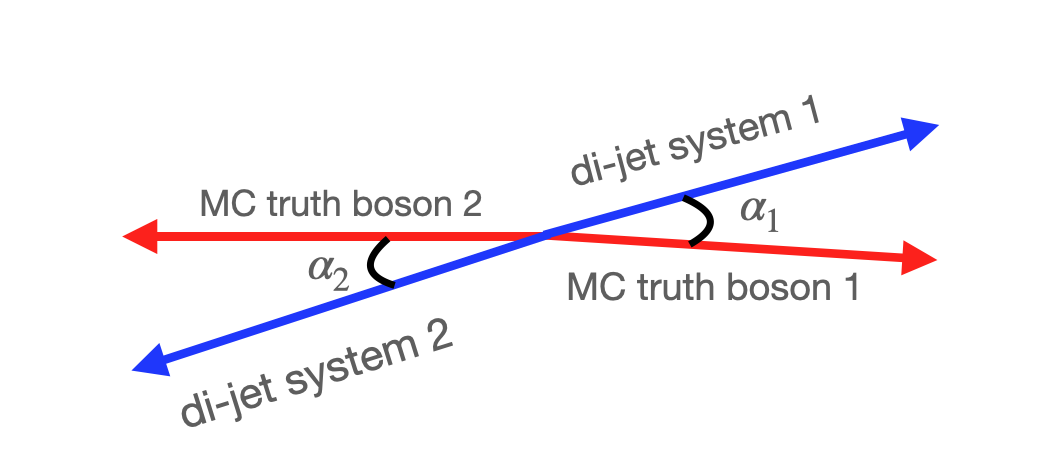}
\caption{\label{fig:alphaDis} The definition of $\alpha_1$ and $\alpha_2$.}
\end{figure}

The performance of the CSI can be evaluated by the angles between the reconstructed bosons and the MC truth bosons, $\alpha_1$ and $\alpha_2$, as shown in figure~\ref{fig:alphaDis}.
Since the CSI evaluator used in this paper uses the MC truth information, it is only a demonstrator to illustrate the importance of an excellent CSI reconstruction.

\begin{figure}[tbp]
\centering 
\includegraphics[width=.45\textwidth]{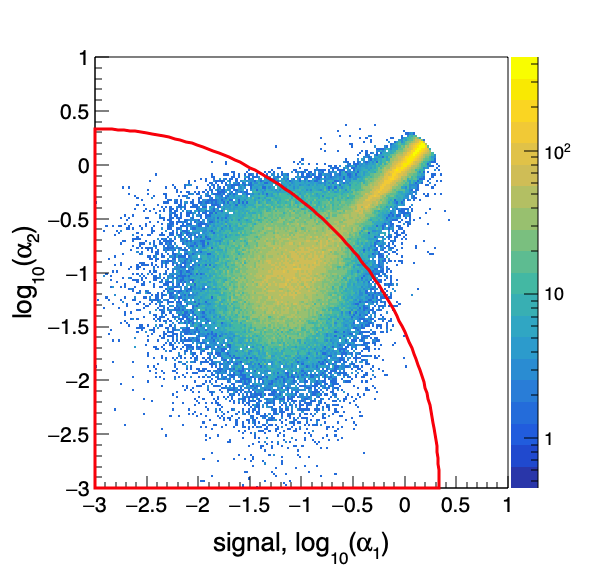}
\hfill
\includegraphics[width=.45\textwidth]{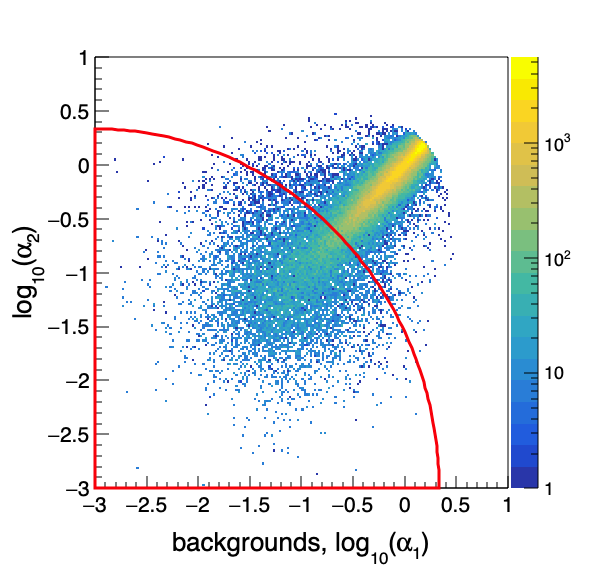}
\caption{\label{alphavs} After the whole event selection in table~\ref{tab:4}, the distributions of $log_{10}(\alpha_1)$ versus $log_{10}(\alpha_2)$ for signal and background are shown in the left and right plots, respectively.}
\end{figure}

\begin{figure}[tbp]
\centering 
\includegraphics[width=.45\textwidth]{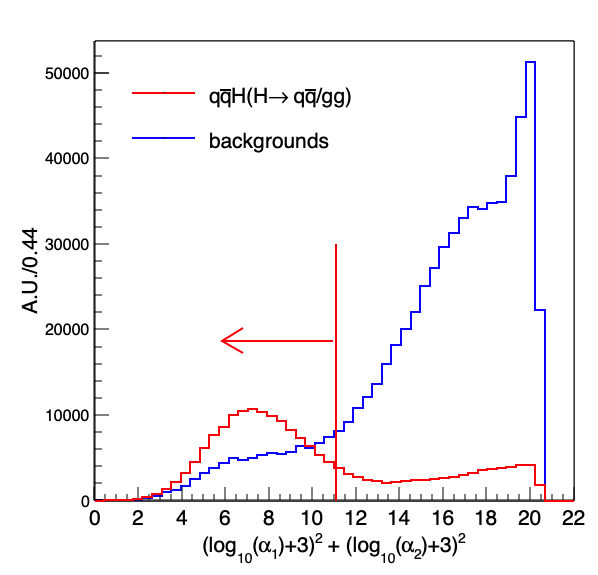}
\hfill
\includegraphics[width=.45\textwidth]{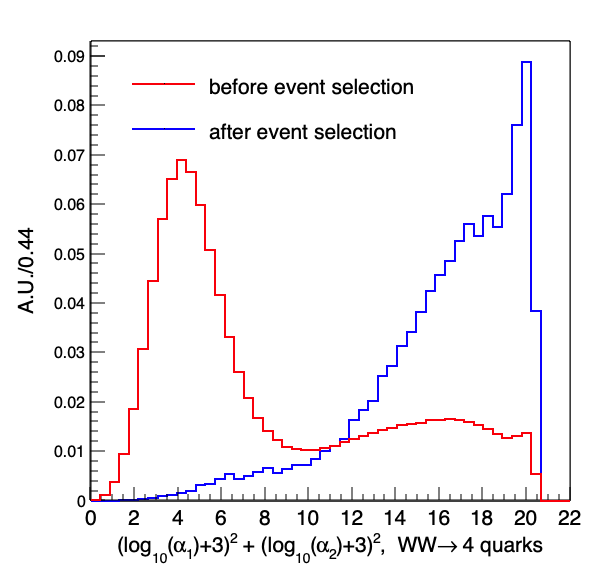}
\caption{\label{fig:14} The distributions of $(log_{10}(\alpha_1)+3)^2 + (log_{10}(\alpha_2)+3)^2$. The left plot corresponds to the signal and backgrounds after the whole event selection in table~\ref{tab:4}. The right plot corresponds to the $e^+e^-\to W^+W^-\to 4\ quarks$ before and after the whole event selection in table~\ref{tab:4} to illustrate that the event selection process was able to strongly suppress the backgrounds with good CSI performance.}
\end{figure}

After the entire event selection in table~\ref{tab:4}, the distributions of $log_{10}(\alpha_1)$ versus $log_{10}(\alpha_2)$ are shown in figure~\ref{alphavs}.
The circle $(log_{10}(\alpha_1)+3)^2 + (log_{10}(\alpha_2)+3)^2 = 11.11$ can improve the signal-to-background ratio.
The distribution of $(log_{10}(\alpha_1)+3)^2 + (log_{10}(\alpha_2)+3)^2$ for the signal and backgrounds is shown in figure~\ref{fig:14}, which shows that most  backgrounds have relatively poor CSI performance compared to the signal events.
This is because the backgrounds with good CSI performance were strongly suppressed by the event selection.
For example, the right plot in figure~\ref{fig:14} shows the distributions of $(log_{10}(\alpha_1)+3)^2 + (log_{10}(\alpha_2)+3)^2$ for the samples of $e^+e^-\to W^+W^-\to 4\ quarks$, where the red line corresponds to all samples, while the blue line corresponds to that after event selection. 
To illustrate the performance of the CSI, each of these two distributions is normalized to a unit area.
We can see that only the backgrounds with poor CSI performance passed the event selection process.
An ideal CSI performance evaluator such as the quantity shown in the left plot of figure~\ref{fig:14} shows a potential to improve the relative accuracy of $q\bar{q}H(H\to b\bar{b}/c\bar{c}/gg)$ signal strength by 6\%/77\%/90\%.
This motivates future developments aimed at improving the CSI reconstruction or developing a performance estimator based on reconstructed quantities.

\subsection{Possible improvements to the flavor tagging}
\label{sec:OptFLV}

Subsections~\ref{sec:OptvvH} and~\ref{sec:OptqqH} quantify the impact of flavor tagging performance on objective measurement, which strongly promotes better flavor tagging performance.
Better flavor tagging performance can be pursued by optimizing the vertex detector and  developing advanced reconstruction algorithms.
The CEPC vertex detector is designed as a barrel-shaped structure with three concentric cylinders of double-sided layers, whose parameters are listed in table~\ref{vtx}.
The main features of the vertex detector are a single-point resolution of the first layer of better than $3\,\mu m$, a material budget of less than 0.15\% X$_0$ per layer, and the location of the first layer near the beam pipe with a radius of $16\,$mm.
The flavor tagging algorithm used in this analysis is implemented in the LCFIPlus package and is based on a GBDT.

\begin{table}[tbp]
\centering
\begin{tabular}{|cccc|}
\hline
                    &   R (mm)      & sigle-point resolution ($\mu m$)     & material budget    \\
Layer 1             &   16          & 2.8                              & 0.15\%/X$_0$    \\
Layer 2             &  18           & 6                              & 0.15\%/X$_0$     \\
Layer 3             & 37            & 4                               & 0.15\%/X$_0$   \\
Layer 4             & 39            & 4                               & 0.15\%/X$_0$   \\
Layer 5             &  58           &  4                            & 0.15\%/X$_0$          \\
Layer 6             &  60           &  4                            & 0.15\%/X$_0$          \\
\hline
\end{tabular}
\caption{\label{vtx} The baseline design parameters of the CEPC vertex system.}
\end{table}

\begin{figure}[tbp]
\centering
\includegraphics[width=0.45\textwidth]{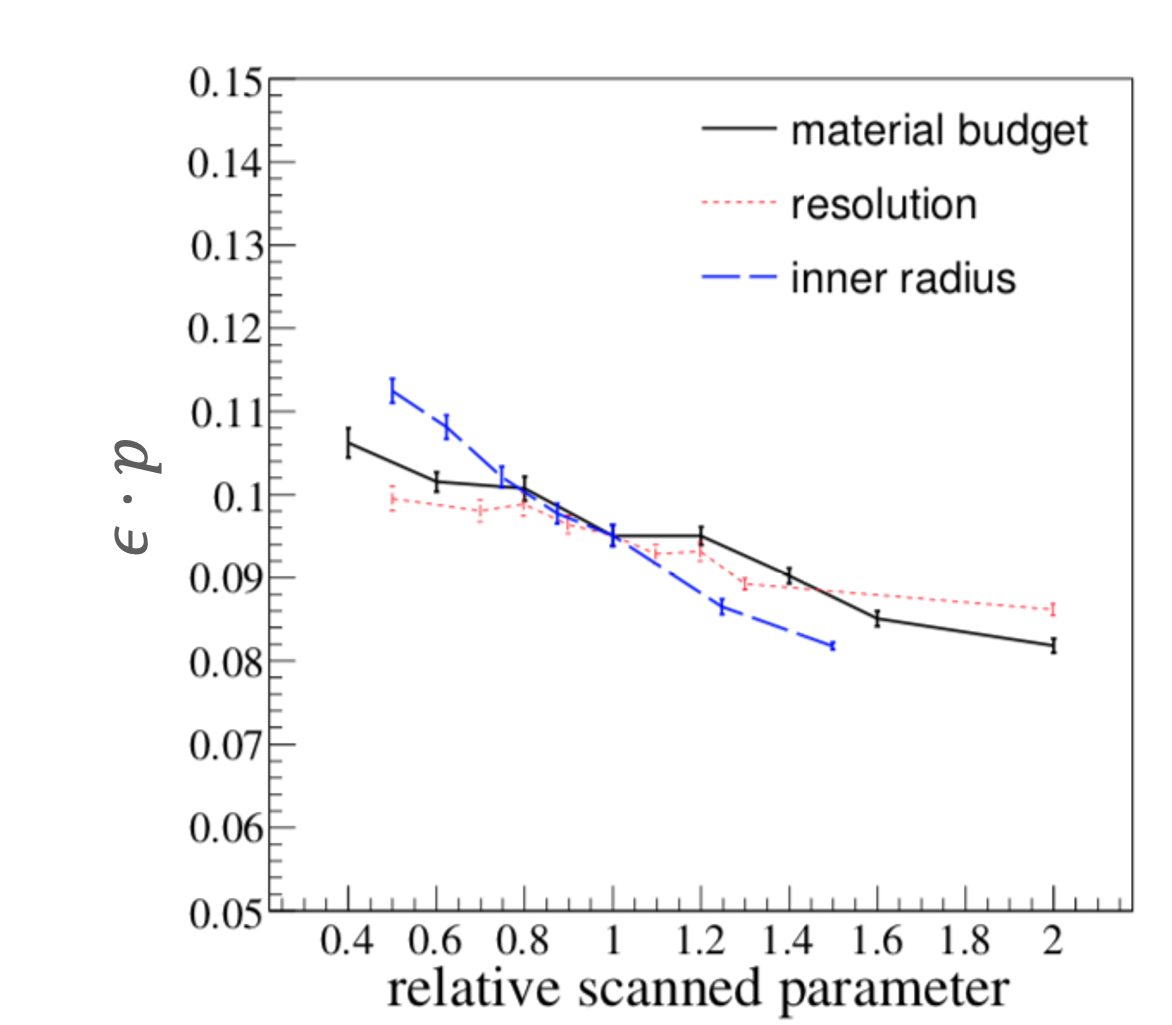}
\caption{The c-tagging performance with a parameter scan on the basis of the CEPC baseline. 
The y-axis represents the efficiency times purity of c-jets tagged in $\nu\nu H, H\to b\bar{b}, c\bar{c}, gg$ samples. 
The x-axis represents the relative difference of vertex detector parameters to the CEPC baseline.}
\label{zhigang}
\end{figure}

For the optimization of the vertex detector, a previous exercise~\cite{VTX} quantifies the correlation between flavor tagging performance and relevant detector properties, including inner radius,  material budget, and spatial resolution of the vertex system, as shown in figure~\ref{zhigang}.
This exercise shows that significant improvements can be achieved if the above properties can be reduced compared to the baseline design.
Considering an optimal and a conservative scenario, as proposed by ref.~\cite{VTX}, where the number of these three parameters is 0.5/2 times compared to the baseline design (the geometry we used in this simulation), the $Tr_{mig}$ will be changed from 2.35 to 2.54/2.16 in the $\nu\bar{\nu}H$ channel, as shown in figure~\ref{fig:11}, and from 2.12 to 2.31/1.93 in the $q\bar{q}H$ channel, as shown in figure~\ref{fig:12}.
The details can be found in appendix~\ref{vtx_FT}.

\section{Discussion of systematic uncertainties}
\label{sec:syst}

The systematic uncertainties relevant to the analyzes presented in this manuscript originate from many sources, including the measurement of the integrated luminosity, the jet energy scale, the track momentum resolution, the reconstructed invariant mass and visible energy of hadronic systems, the flavor tagging performance, the jet configuration and the CSI, some of which are discussed below.

\begin{itemize}
\item According to the CEPC conceptual design report~\cite{CEPCStudyGroup:2018ghi}, the integrated luminosity is required to be measured with a relative accuracy of $10^{-3}$ for CEPC Higgs operation and $10^{-4}$ for Z-pole operation.
The systematic uncertainties caused by the uncertainty of the integrated luminosity are therefore negligible for the $H\to c\bar{c}/gg$ signal strength measurement, while it reaches a comparable level for the $H\to b\bar{b}$ measurement with an integrated luminosity of $20\,ab^{-1}$~\cite{CEPCPhysicsStudyGroup:2022uwl, Gao:2022lew}. 
The accuracy requirements for the integrated luminosity measurement should be tightened to cope with the accuracy of the $H\to b\bar{b}$ signal strength measurement at the $10^{-4}$ level, i.e. to an accuracy of $5\times 10^{-4}$.
Since the statistics of the main physics processes of luminosity monitoring, i.e. the small-angle Bhabha and di-photon events, scale with the integrated luminosity, this goal shall in principle be possible, while a very precise control of Lumi-Cal installation, calibration, and monitoring would be essential.
\item The event selection relies on the reconstruction of the hadronic system, in particular its momentum and energy.
Therefore, understanding the jet energy scale/resolution is crucial to control the systematic uncertainty.
The analysis in ref.~\cite{Lai:2021rko} shows that the jet energy scale can be controlled within 0.5\% at the baseline CEPC detector, leading to an uncertainty of the order of $10^{-4}$ on the selection efficiency mainly due to the cut on the recoil mass.
These uncertainties can be further reduced by data-driven methods, i.e. by reconstructing the differential jet energy scale in-situ using semi-leptonic WW and ZZ events as well as $e^+e^-\to q\bar{q}$ events.
Similarly, the track momentum scale and the photon energy scale can in principle be calibrated using candles of narrow resonances such as $K_S^{0}$~\cite{tai}, $\Lambda$, $J/\psi$, and $\pi^0$, since these physics objects are abundant in hadronic events.
The in-situ calibration can be applied to control the relevant systematic errors.
\begin{figure}[tbp]
\centering
\includegraphics[width=0.55\textwidth]{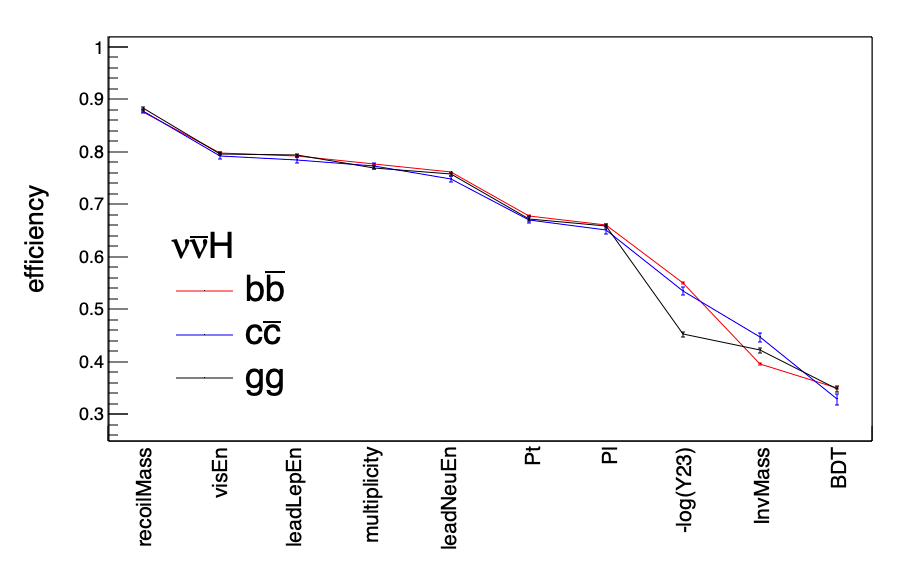}
\caption{The selection efficiency of $\nu\bar{\nu}H(H\to b\bar{b}/c\bar{c}/gg)$ for each cut variable. }
\label{eff}
\end{figure}
\item The total energy and momentum of the hadronically decaying Higgs boson significantly depends on its decay modes, since the heavy flavor quarks can decay semi-leptonically.
The semi-leptonic decays of b/c quarks generate neutrinos, leading to a significant deformation of the reconstructed invariant mass for $H\to b\bar{b}/c\bar{c}$ events, which in turn results in a flavor dependent efficiency in the event selection chain, as shown in figure~\ref{eff}.
To accurately calculate the corresponding efficiencies for different flavors, it is essential to control the shape of the relevant distribution, which can be done by several methods.
First, the charged lepton from the semi-leptonic decay can be identified efficiently~\cite{Yu:2021pxc}.
Second, using physics events with a pair of b-jets (i.e. $e^+e^-\to q\bar{q}$ events, $\nu\bar{\nu}Z$ events, and Z-pole events) and a restrictive selection on the b-likeness of a jet, we can obtain highly pure and inclusive b-jet samples, since the decay modes of both b-jets are in principle independent.
Third, the $\ell^+\ell^-H$ channel with lepton identification inside the jet also offers the possibility to control the b-invariant mass.
To conclude, there are multiple ways to mitigate this effect.
\item A much more subtle systematic for the $H\to gg$ measurements arises from the fact that the gluon jet has a different configuration compared to the quark jet.
As shown in figure~\ref{eff}, the Y23 cut leads to an efficiency of 83\%/82\%/68\% for the $H\to b\bar{b}/c\bar{c}/gg$ selection.
For the $q\bar{q}H$ channel, as shown in figure~\ref{qqheff}, the thrust cut results in an efficiency of 87\%/87\%/93\% for the $H\to b\bar{b}/c\bar{c}/gg$ selection.
The cut variables related to the jet configuration in the $q\bar{q}H$ channel also include -log(Y34) and the angle of the two jets from the heavy/light di-jet system.
Therefore, it is essential to understand and calibrate the spatial configurations of the jets.
This requirement can be addressed using sophisticated QCD calculations and its comparison with low pile-up LHC data as well as three-jet events in CEPC Z-pole operation.
\begin{figure}[tbp]
\centering
\includegraphics[width=0.55\textwidth]{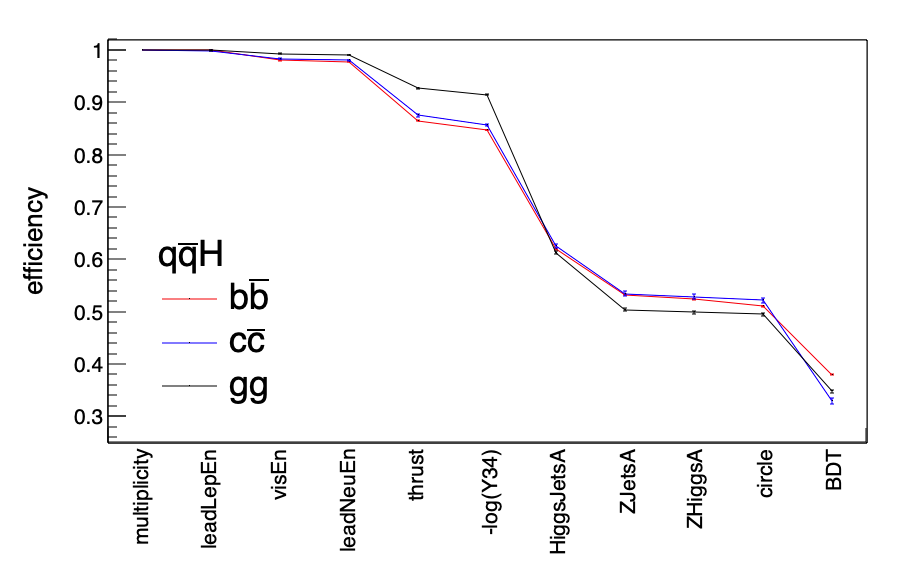}
\caption{The selection efficiency of $q\bar{q}H(H\to b\bar{b}/c\bar{c}/gg)$ for each cut variable. }
\label{qqheff}
\end{figure}
\item The systematic uncertainties caused by flavor tagging performance can be characterized by the uncertainties in the flavor tagging performance matrix, i.e. the difference between the actual flavor tagging performance matrix and that obtained from simulation.
Using the abundant hadronic and semi-leptonic events at CEPC, as well as the prior knowledge of the branching ratios of the W and Z boson decays into different quark flavors, we can derive the flavor-tagging performance matrix using a data-driven method.
The statistics of relevant hadronic events, e.g. WW, ZZ, ISR-return-Z processes in CEPC Higgs runs, is 2-3 orders of magnitude higher than that of the Higgs signal.
In addition, the CEPC is expected to acquire several $10^{12}$ hadronic Z events at its Z-pole operation, 6 orders of magnitude larger than the expected number of Higgs events.
These samples, especially the semi-leptonic WW events and the hadronic Z events, can be controlled with very high purity.
Therefore, data-driven methods could control the systematic uncertainties caused by the flavor tagging performance to a negligible level if the following two conditions are met.
The first is that the detector is well understood and maintains stable during the physics data acquisition, which requires dedicated performance analysis and detector stability analysis.
The second is that the relative difference between gluon jets and quark jets on the behavior of flavor tagging is well controlled, which requires dedicated QCD studies and performance studies.
\item CSI is essential for the $q\bar{q}H$ analyzes.
Not only because the ideal CSI can greatly improve final accuracy, but also because the CSI induces a flavor-dependency in the event selection efficiency.
The circular cut (the tenth row in table~\ref{tab:4}) $(M_{heavy} - 125)^2 + (M_{light} - 91)^2 <= 29^2$ based on the information from the CSI has an efficiency of 97\%/99\%/99\% for $H\to b\bar{b}/c\bar{c}/gg$.
In this manuscript, the CSI is based on the jet clustering and jet matching procedures.
To investigate the systematic uncertainty caused by the CSI, we replace the Durham algorithm with Valencia algorithm~\cite{VLC} and obtain efficiencies of 97\%/99\%/99\% for $H\to b\bar{b}/c\bar{c}/gg$.
The difference between the Durham and the Valencia is smaller than the MC statistical uncertainty.
But the relative difference in event selection efficiency between the different flavors is in the percentage level and has to be controlled in a further step.
\end{itemize}

To conclude, we categorize the leading systematic uncertainties in these analyzes into three groups.
The first group are those that are significantly smaller than the statistical uncertainties, including the reconstructed energy/momentum scale of the physics objects.
The second group are those comparable to the statistical uncertainty, especially the integrated luminosity.
The third group are those that can be significantly larger than the statistical uncertainty, including CSI and the jet configuration.
A full quantification of the systematic uncertainty is beyond the scope of this paper and awaits real data and new methods.

\section{Conclusion}

We estimate the anticipated accuracy for the $H\to b\bar{b}/c\bar{c}/gg$ measurements at CEPC with its nominal luminosity of $5.6\,ab^{-1}$ corresponding to the CEPC CDR~\cite{CEPCStudyGroup:2018ghi} and $20\,ab^{-1}$ as proposed for Snowmass 2021~\cite{Gao:2022lew}.
Using the CEPC CDR baseline detector, we simulated MC samples corresponding to the CEPC Higgs operation and combined the accuracies obtained in the $\ell^+\ell^-H$, $\nu\bar{\nu} H$, and $q\bar{q}H$ channels.
We conclude that the signal strength of $H\to b\bar{b}/c\bar{c}/gg$ can be measured with a statistical uncertainty of 0.27\%/4.03\%/1.56\% and 0.14\%/2.13\%/0.82\%, corresponding to integrated luminosities of $5.6\,ab^{-1}$ and $20\,ab^{-1}$, respectively.
In addition, we discuss the relevant systematic uncertainties, critical performance, and vertex system optimization.
We also identify several critical topics that should be studied in detail in the future.

We found that the systematic uncertainty caused by the integrated luminosity is comparable to the statistical uncertainty of the $H\to b\bar{b}$ signal strength measurement, that the systematic uncertainty caused by CSI and the jet configuration can be much larger than the statistical uncertainty, and that there are multiple ways to control the systematic uncertainties caused by reconstructed hadronic systems.
Data-driven methods are expected to control the systematic uncertainty.
Moreover, the complicated patterns and the deviations of the branching ratios of the Z-boson decay from the naive expectations in figure~\ref{fig:qqHcate} show that flavor tagging would lead to serious systematic uncertainties that need to be controlled based on a much better understanding of fragmentation, hadronization, and gluon splitting.

The dependence of the measurement accuracy on critical detector performance aspects, specifically flavor tagging and CSI, has been analyzed. 
Compared with the flavor tagging performance of the baseline CEPC detector, perfect flavor tagging performance could improve the relative accuracy of the $H\to b\bar{b}/c\bar{c}/gg$ signal strength by 35\%/122\%/181\% in the $q\bar{q}H$ channel and 2\%/63\%/13\% in the $\nu\bar{\nu} H$ channel.
An ideal CSI or a reliable evaluator for CSI performance can significantly improve the physics reach, which motivates us to pay more attention to CSI.

From our analysis of relevant performance and systematic uncertainties, we conclude that the critical detector and reconstruction performance of flavor tagging and CSI has a very strong impact on the anticipated precision.
Therefore, we would like to encourage the design and optimization of the vertex system towards better precision, smaller inner radius, and lower material budget.
The development of advanced reconstruction algorithms, probably synchronized QCD studies, to achieve a better CSI performance.
The per-mille accuracy also places high demands on systematic control, especially on the stability of detector operation.
We also note a significant difference in spatial configuration between the quark and gluon jets, which can lead to significant systematic uncertainties.
Dedicated studies of the theoretical calculation of QCD and comparison with the available data are therefore crucial to control these uncertainties.

\appendix
\section{Cross section, expected and simulated event number, and scaling factor}
\label{appendix}

The table~\ref{tab:sample} lists the cross section, the number of expected events, the number of simulated events, and the scaling factors used in this analysis.
The scaling factor is defined as the simulated statistic divided by the expected statistic.
The single-Z process consists of an electron-positron pair and an on-shell Z boson in the final state.
The single-W process consists of a $e^{\pm}$ together with its neutrino and an on-shell W boson in the final state.
The ZZ and WW processes consist of two on-shell bosons decaying into four fermions. 
The mixed process consists of two mutually charge-conjugated pairs in the final state, which could be from either the virtual WW or the ZZ.

\begin{table}[tbp]
\centering
\begin{footnotesize}
\begin{tabular}{|cccccc|}
\hline
\multirow{2}{*}{name}    & \multirow{2}{*}{channel}                   &X-section            & expected       &simulated               &scaling factor            \\
                                      &                                                           &  $fb$        & million          & k                & \% \\
\hline
\multirow{5}{*}{ZH}   & $Z\to \nu\nu, Higgs\ inclusive\ decay$              &46.29     &0.26               &217 &83 \\
                                 & $Z\to e^+e^-, Higgs\ inclusive\ decay$              &7.04      &0.04                &87 &221 \\
                                 & $Z\to \mu^+\mu^-, Higgs\ inclusive\ decay$      &6.77     &0.04                 &72 &191 \\
                                 & $Z\to \tau^+\tau^-, Higgs\ inclusive\ decay$      &6.75     &0.04                &82 &217 \\
                                 & $Z\to q\bar{q}, Higgs\ inclusive\ decay$            &136.81 &0.77                &566 &74 \\
\hline
\multirow{15}{*}{ZZ} & $Z\to c\bar{c}, Z\to d\bar{d}/b\bar{b}$              &98.97     &0.55         &123 &22 \\
                                 & $ZZ\to 4\ down\ quarks$                                  &233.46  &$1.31        $ &288 &22 \\
                                 & $ZZ\to 4\ up\ quarks$                                       &85.68    &0.48 &102 &21 \\
                                 & $Z\to u\bar{u}, Z\to s\bar{s}/b\bar{b}$              & 98.56   &0.55 &120 &22 \\
                                 & $Z\to \mu^+\mu^-, Z\to \mu^+\mu^-$               & 15.56   &0.09 &23 & 26\\
                                 & $Z\to \tau^+\tau^-, Z\to \tau^+\tau^-$               & 4.61     &0.03 &25 &97 \\
                                 & $Z\to \mu^+\mu^-, Z\to \nu_{\tau}\nu_{\tau}$   &  19.38  &0.11 &24 &22 \\
                                 & $Z\to \tau^+\tau^-, Z\to \mu^+\mu^-$               & 18.65   &0.10 &22 &21 \\
                                 & $Z\to \tau^+\tau^-, Z\to \nu{\tau}\nu{\tau}$       & 9.61     &0.05 &25 &46 \\
                                 & $Z\to \mu^+\mu^-, Z\to down\ quarks$            & 136.14  &0.76 &644 &85 \\
                                 & $Z\to \mu^+\mu^-, Z\to up\ quarks$                 & 87.39   &0.49 &110 &23 \\
                                 & $Z\to \nu\nu, Z\to down\ quarks$                     & 139.71 &0.78 &175 &22 \\
                                 & $Z\to \nu\nu, Z\to up\ quarks$                          & 84.38   &0.47 &105 &22 \\
                                 & $Z\to \tau^+\tau^-, Z\to down\ quarks$             &67.31    &0.38&312 & 83\\
                                 & $Z\to \tau^+\tau^-, Z\to up\ quarks$                  & 41.56   &0.23 &193 &83 \\
\hline
\multirow{7}{*}{WW} & $ccbs$                                                          &5.89            & 0.03      & 24   &75 \\
                                 & $ccds$                                                          &170.18        & 0.95      & 203 &21 \\
                                 & $cusd$                                                          &3478.89      & 19.5      &2668 & 14\\
                                 & $uusd$                                                           &170.45       & 0.95      &194 &4.92 \\
                                 & $WW\to 4-lepton$                                          & 403.66      & 2.26      &488 &20 \\
                                 & $W\to mu\nu_{\mu}, W\to qq$                        &2423.43     & 13.6      &9215 & 68\\
                                 & $W\to tau\nu_{\tau}, W\to qq$                         & 2423.56   & 13.6     &2745 &20 \\
\hline
\multirow{3}{*}{SW}  & $e\nu_{e},W\to \mu\nu_{\mu}$ &436.70       & 2.44 &538                          &22 \\
                                 & $e\nu_{e},W\to \tau\nu_{\tau}$ &435.93       & 2.44 &535                          &22 \\
                                 & $e\nu_{e},W\to qq$                  &2612.62      & 14.63 &9233                      &63 \\
\hline
\multirow{10}{*}{SZ}  & $e^+e^-, Z\to e^+e^-$                   &78.49    &0.44 &97                              &22 \\
                                 & $e^+e^-, Z\to \mu^+\mu^-$           &845.81   &4.74 &520                            &11 \\
                                 & $e^+e^-, Z\to \nu\nu$                    &28.94     &0.16 &36                              &22 \\
                                 & $e^+e^-, Z\to \tau^+\tau^-$           &147.28   &0.82 &180                             &22 \\
                                 & $e^+e^-, Z\to down\ quarks$        &125.83   &0.70 &153                             &22 \\
                                 & $e^+e^-, Z\to up\ quarks$             &190.21   &$1.06$ &231                         &22 \\
                                 & $\nu^+\nu^-, Z\to \mu^+\mu^-$     &43.42     &0.24 &37                               &15 \\
                                 & $\nu^+\nu^-, Z\to \tau^+\tau^-$     &14.57     &0.08 &22                               &27 \\
                                 & $\nu^+\nu^-, Z\to down\ quarks$  &90.03     &0.50 &90                               &18 \\
                                 & $\nu^+\nu^-, Z\to up\ quarks$       &55.59     &0.31 &71                               &23 \\
\hline
\multirow{5}{*}{mix}  & $ZZ/WW\to\mu\mu\nu_{\mu}\nu_{\mu} $           &221.10         &1.24   &263 &21 \\
                                 & $ZZ/WW\to \tau\tau\nu_{\tau}\nu_{\tau}$           &211.18         &1.18   &262 &83 \\
                                 & $ZZ/WW\to ccss$                                               &1607.55      &9.00    &1966 &22 \\
                                 & $ZZ/WW\to uudd$                                              &1610.32      &9.02    &1604 &18 \\
                                 & $SW/SZ\to ee\nu_{e}\nu_{e}$                                            &249.48         &1.40   &307 &22 \\
\hline
\multirow{3}{*}{2f}      &  $e^+e^-$                &24770.90                    &138.72      &314                  &0.23          \\
                                 &  $\mu^+\mu^-$        &5332.71                      &29.86      &278                    &0.93       \\
                                 &  $\tau^+\tau^-$        &4752.89                      &26.62      &746                    &2.80      \\
                                &   $q\bar{q}$              &54106.86                    &303.00      &7437                &2.45       \\
\hline
\multicolumn{2}{|c}{$\gamma\gamma\to hadrons$}           & 87670                          & 490.95           & 1366     &  0.28 \\
\hline
\end{tabular}
\end{footnotesize}
\caption{\label{tab:sample} The cross section, expected event number, simulated event number, and scaling factor of the signal and various backgrounds.}
\end{table}

\section{Dependence of flavor tagging performance on vertex detector design}
\label{vtx_FT}

The relationship between c-tagging efficiency times purity ($\epsilon\cdot p$) and $Tr_{mig}$ in the $q\bar{q}H$ channel is fitted with an empirical formula as $Tr_{mig} = 1.11\cdot log_{10}(\epsilon \cdot p) + 3.23$.
The value of $\epsilon \cdot p$ ranges from 0.02 to 0.8.
Considering the relationship between $\epsilon\cdot p$ and vertex detector parameters shown in figure~\ref{zhigang}, the empirical formula for $Tr_{mig}$ and vertex detector parameters is 
\begin{equation}
\label{eq:x_cFo}
Tr_{mig} = 2.12 +0.05 \cdot log_{2}\frac{R_{material}^0}{R_{material}} +0.04 \cdot log_{2}\frac{R_{resolution}^0}{R_{resolution}} +0.10 \cdot log_{2}\frac{R_{radius}^0}{R_{radius}},
\end{equation}
where $R_{material}^0$ is the default material budget and $R_{material}$ is the modified material budget, and similarly for the other parameters.
In the $\nu\bar{\nu}H$ channel, the relationship between c-tagging efficiency times purity and $Tr_{mig}$ is fitted to an empirical formula as $Tr_{mig} = 1.12\cdot log_{10}(\epsilon \cdot p) + 3.28$.
The value of $\epsilon \cdot p$ ranges from 0.02 to 0.8.
The empirical formula for $Tr_{mig}$ and the parameters of the vertex detector is 
\begin{equation}
\label{eq:x_cFo}
Tr_{mig} = 2.35 +0.05 \cdot log_{2}\frac{R_{material}^0}{R_{material}} +0.04 \cdot log_{2}\frac{R_{resolution}^0}{R_{resolution}} +0.10 \cdot log_{2}\frac{R_{radius}^0}{R_{radius}}.
\end{equation}

\acknowledgments

We thank Chengdong FU, Gang LI, and Xianghu ZHAO for providing the simulation tools and samples.
We thank Hao Liang, Dan Yu, and Yudong Wang for useful discussions.
This project is supported by the International Partnership Program of Chinese Academy of Sciences (Grant No. 113111KYSB20190030), the Innovative Scientific Program of Institute of High Energy Physics.

\end{document}